\documentclass[12pt,authoryear]{elsarticle}

\usepackage{amssymb}

\usepackage{placeins}
\usepackage{caption}
\usepackage{subcaption}
\usepackage{graphicx}
\usepackage[lofdepth,lotdepth]{subfig}
\usepackage{comment}
\usepackage[f]{esvect}
\usepackage{url}
\usepackage{listings}
\journal{Journal of Theoretical Biology}

\begin{document}

\begin{frontmatter}

\title{Use of 3D chaos game representation to quantify DNA sequence similarity with applications for hierarchical clustering}

\author[label1]{Stephanie Young\corref{cor1}}
\cortext[cor1]{Corresponding author.}
\ead{syoung2@sdsu.edu}
\author[label1,label3]{J\'er\^ome Gilles}

\affiliation[label1]{organization={Computational Science Research Center, San Diego State University},
            addressline={5500 Campanile Dr}, 
            city={San Diego},
            postcode={92182}, 
            state={CA},
            country={USA}}

\affiliation[label3]{organization={Department of Mathematics and Statistics, San Diego State University},
            addressline={5500 Campanile Dr}, 
            city={San Diego},
            postcode={92182}, 
            state={CA},
            country={USA}}

\begin{abstract}
A 3D chaos game is shown to be a useful way for encoding DNA sequences. Since matching subsequences in DNA converge in space in 3D chaos game encoding, a DNA sequence's 3D chaos game representation can be used to compare DNA sequences without prior alignment and without truncating or padding any of the 
sequences. Two proposed methods inspired by shape-similarity comparison techniques show that this form of encoding can perform as well as alignment-based techniques for building phylogenetic trees. The first method uses the volume overlap of intersecting spheres and the second uses shape signatures by summarizing the coordinates, oriented angles, and oriented distances of the 3D chaos game trajectory. The methods are tested using: 1) the first exon of the beta-globin gene for 11 species, 2) mitochondrial DNA from four groups of primates, and 3) a set of synthetic DNA sequences. Simulations show that the proposed methods produce distances that reflect the number of mutation events; additionally, on average, distances resulting from deletion mutations are comparable to those produced by substitution mutations. 

\end{abstract}




\end{frontmatter}

\section{Introduction}
\label{intro}

Quantifying DNA sequence similarity is a challenge in bioinformatics that has been amplified in recent years by the development of high-throughput sequencing technologies. The ever-expanding volume of DNA sequence data being made available has stimulated research into DNA representations that can facilitate efficient and accurate comparisons, classifications, and clustering of sequence data.

Traditionally, DNA sequence comparisons have involved an alignment step to enable nucleotide-by-nucleotide comparisons of sequences. The evolutionary distance between any pair of DNA sequences is typically calculated as a function of the number of mismatching nucleotides. Simpler distance metrics are based on the count of differences between aligned sequences. More sophisticated metrics, however, utilize parameters that account for differing types and rates of mutations, thereby enabling them to contribute variably to the estimated distances. These sorts of alignment-dependent comparisons are not only intuitive but also advantageous as their associated distance metrics can be grounded in explicit models of sequence evolution.

However, alignment-based approaches have several drawbacks. First, the choice of sequences for DNA sequence alignment can affect the quality and robustness of the alignment and, hence, the inferred relationships from post-alignment analysis. Sequences need to be sufficiently different that they can reveal meaningful information about their evolutionary history. Yet sequences cannot be so different that alignment is too difficult to reliably perform. Second, sequence-alignment may not be appropriate for sequences that have undergone numerous recombination mutations as the alignment may not accurately reflect evolutionary distance. Finally, sequence alignment can be very resource intensive, especially as the number of sequences increases.

Due to these challenges associated with sequence alignment, numerous alignment-free methods have been developed to handle the deluge of sequence data. The most widely used alignment-free methods for comparing DNA sequences are k-mer based methods which focus on enumerating the occurrences of short subsequences (k-mers) of a defined length $k$. These alignment-free alternatives tend to be robust against recombination events and are computationally efficient and highly scalable. However, there are a number of drawbacks from alignment-free methods. These methods tend to lack the flexibility of alignment-based methods to incorporate information from evolutionary models and to allow customizable penalties for different mutation types. They can be sensitive to parameterization and not all distance metrics translate into reliable downstream analyses as phylogenetic reconstruction based on alignment-free methods often vary in quality. The benchmarking of these techniques often assesses how well the method reconstructs a known, ground-truth phylogeny based on the sequences using metrics such as the Robinson-Foulds distance measure to quantify difference between the ground-truth phylogeny and one built from the alignment-free distance metrics. However, this approach is largely indifferent to identifying or knowing the very evolutionary processes and mutations that give rise to the sequences examined. 

In addition to k-mers, a number of proposed alignment-free methods involve clever recoding of the biological sequence data such as DNA or protein sequences such that the sequences can be analyzed in the framework of certain computational techniques that are fast and/or powerful. For example, \citet{Yin:2014Fourier} and \citet{Yin:2014Ramanujan} recode the DNA into a set of binary indicator sequences so that they can perform Fourier and Ramanujan-Fourier transforms on the data; similarly, \citet{Rockwood} recode DNA sequences into complex-valued vectors to leverage the fast Fourier transform and to calculate cross correlations. Other authors such as \citet{SENSE} and \citet{NeuroSEED} utilize neural network encoder models to map sequences to an embedding space with the goal of producing an embedding distance that is representative of the edit or evolutionary distance between the sequences. 

However, often glossed over in such proposed alignment-free approaches to whole-sequence comparisons is that there is often a need to address sequences of different lengths. Some authors pad the sequences with zero values to force all sequences to be of equal length \citep{SENSE, Vinga:2012}; another alternative approach is to stretch sequences in the frequency domain after a fast Fourier transform prior to any distance calculation \citep{Yin:2014Fourier, Yin:2014Ramanujan, Hoang:2016}. But limitations as well as distortions introduced by these approaches are often not explicitly discussed. For example, the choice of zero-padding has the potential to strongly distort or affect a neural network's ability to correctly classify a sequence type \citep{zeropadding} and stretching sequences in the frequency domain results in distances induced by deletions/insertions that are substantially greater than those induced by substitution mutations. In fact, in the paper by \citet{Yin:2014Fourier}, a single base-pair deletion creates a distance from the original sequence that is more than 1000 times that created by a single base-pair substitution. Unless one is interested in heavily penalizing deletion mutations, the method can only perform well under the condition that the evaluated DNA sequences are the same length. Thus, while a number of interesting alignment-free techniques have been developed, many handle unequal sequence lengths poorly and the effect of mutation type on distances on those methods remains not fully characterized. 

\subsection{Chaos Game Representation}

\begin{figure}
   \centering
   \includegraphics[width=0.45\linewidth]{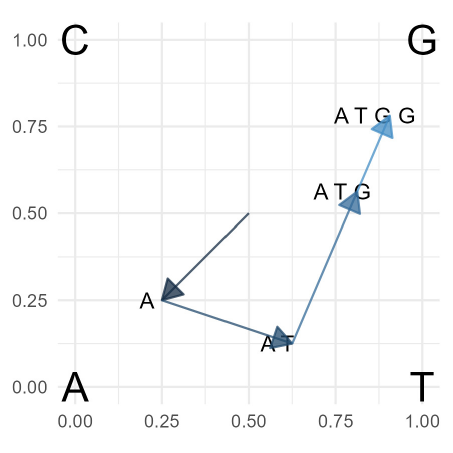}
   \caption{An example of 2D chaos game representation for ATGG.}
   \label{fig:2Dexample}
\end{figure}

The Chaos Game Representation (CGR), especially in 2D, has been frequently utilized as an alignment-free method to capture information about DNA sequences \citep{Vinga:2012, Almeida:2001, Hoang:2016, Huang:3Dchaos}. In 2D CGR, each vertex of the unit square is assigned to one of the four possible nucleotides, e.g., A = (0, 0), C = (0, 1), G = (1, 1), T = (1, 0). The representation starts at the center of the square, at $\left(\frac{1}{2}, \frac{1}{2}\right)$. For each nucleotide in the DNA sequence of interest, the subsequent CGR value is positioned halfway between the previous CGR value and the corresponding corner of the nucleotide. An example of how a 2D CGR is iteratively built for a small subsequence is presented in Figure \ref{fig:2Dexample}.

CGR provides a computationally efficient, lossless representation, wherein all possible sequences have a unique representation. Theoretically, an entire sequence can be recovered from its final CGR coordinate, with the resolution of the final coordinate being the only limiting factor as to how much of the sequence can be recovered. CGR has numerous useful properties: the 2D CGR plane is a scale-independent generalization of Markov chain probability tables \citep{Almeida:2001}; it can be used to identify local alignment points in DNA sequences \citep{Almeida:2001, Joseph:2006}; and the entropic profiles of sequences can be identified through estimations of local densities \citep{Vinga:2007}. 

However, a thorough exploration of the entire trajectory and structure of a chaos game representation to quantify global similarity between entire sequences has been mostly unexamined. A point-by-point comparison of the trajectory is typically restricted to local comparisons because it requires 
sequences be of equal length and/or that the sequences be aligned (similar to traditional, alignment-based sequence comparison methods) prior to comparison. Even in the circumstance where all DNA sequences to be compared are identical in length, the 2D CGR may force some nucleotides to be arbitrarily farther apart in space in a way that has potential downstream consequences. For example, consider using a Euclidean distance to compare sequences of CA, GA, and TA. The trajectory of CA would be $(0.25, 0.75) \rightarrow (0.125, 0.375)$; the trajectory of GA would be $(0.75, 0.75) \rightarrow (0.375, 0.375)$; the trajectory of TA would be $(0.75, 0.25) \rightarrow (0.375, 0.125)$. At every step of the trajectory, CA is closer to GA than it is to TA because the C vertex is positioned closer the G vertex in the 2D-plane; this is despite the fact that both GA and TA are a single nucleotide different from CA. 
So the distance between any two sequences is not only a function of sequence similarity but also depends on the spatial representation of the 2D CGR. As such, the inflated dissimilarity between the CA and TA (or any general pair of sequences) is not grounded in biological reality. 

Many applications of 2D CGR handle unequal sequence lengths using frequency counts (FCGR) at various grid resolutions or using a function of the CGR points to calculate a distance between sequences and find local homology between them \citep{Almeida:2001, Joseph:2006}. From these studies, the use of the 2D CGR to form local-level comparisons of DNA sequences is well established. But its use to determine a global similarity based on point-level comparisons or the entirety of its trajectory has not been fully realized. The method proposed by \citet{Hoang:2016} is one attempt to perform a global similarity comparison by examining the CGR in the frequency domain, but this approach also suffers from the same issues noted in \citet{Yin:2014Fourier} where deletion mutations strongly influence the final distance metrics. 

In this paper, we propose an alignment-free sequence analysis method that utilizes a 3D chaos game encoding to represent DNA sequences and quantifies similarity and/or distance between sequences based on the resulting 3D trajectory by using shape-comparison techniques. Our method requires no sequence padding or stretching that distorts the information embedded. Instead, we focus on a shape-similarity comparison of the overall trajectory of sequences to evaluate their distances. Our method is unique in that distances generated by deletion mutations vs substitution mutations are often comparable in a way that we hope links the estimated distance metric to the number of evolutionary events. 

In the Methods section, we: 1) describe the construction of the 3D CGR, 2) discuss similarity/distance measures used to compare the 3D CGR, 3) discuss the phylogenetic algorithm used to test downstream analysis of CGR-based distances, and 4) discuss the simulations used to evaluate the effect of different mutation types on distance under the proposed methodology. In the Data section, we describe the data we used to test our proposed method. In the Results section, 1) we show how the method performs for phylogenetic analysis, 2) demonstrate the relative stability of estimated distances across a wide range of parameter settings, and 3) through simulations, show how the method performs under different types of mutations. In the Application section, we apply our method to a set of SARS-CoV-2 sequences collected between 2020 and early 2023 to demonstrate its effect on a substantially larger dataset as well as to engage in a criticism of the literature on alignment-based methods and the use of phylogenetic trees as the standard by which we judge their performance. Finally, in the Discussion section, we summarize our findings and discuss the performance of our method relative to other existing alignment-free sequence comparison tools in the AFproject \citep{AFproject} which provides objective benchmarking against known references. 

\section{Methods}
\label{methods}
\subsection{Construction of 3D chaos game representation of DNA sequences}
\label{3DCGR}

\begin{figure}
   \centering
   \includegraphics[width=.65\linewidth]{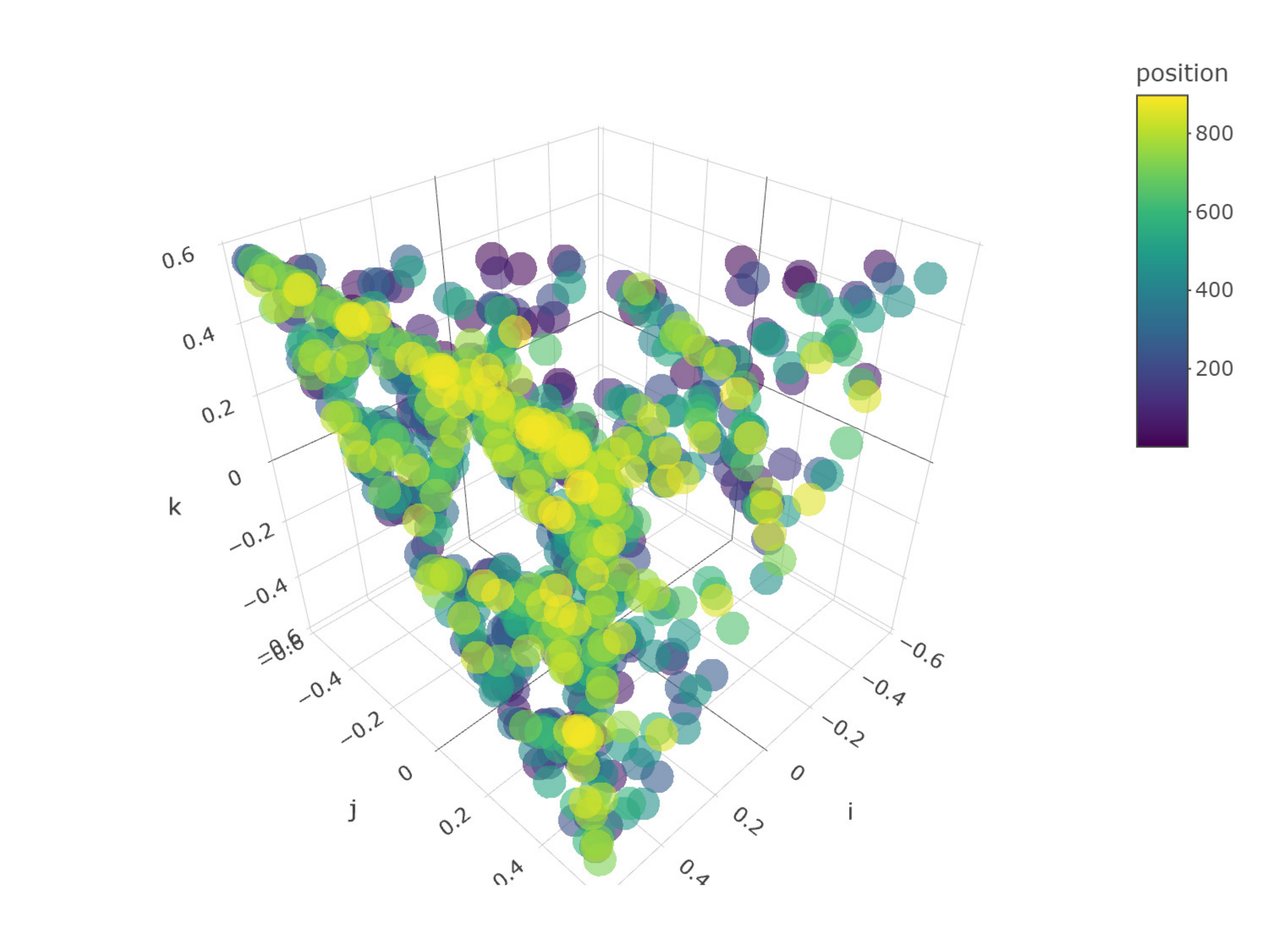}
   \caption{3D chaos game representation of the MT-NADH gene for the Macaca fascicular. Each point represents a nucleotide in the DNA sequence. The color encodes the position of the nucleotide in the sequence with the first element in the sequence coded in purple and the last element in the sequence coded in yellow.}
   \label{fig:nadh1_triangle}
\end{figure}

The proposed 3D CGR forms a Sierpinksi tetrahedron where each of the four possible nucleotides are vertices on the tetrahedron.  The CGR vertices are coded as coordinates in three-dimensional space in the following manner: 
A = $\alpha \left(-1, -1, -1\right)$;
T = $\alpha \left(1, 1, -1\right)$;
C = $\alpha \left(1, -1, 1\right)$;
G = $\alpha \left(-1, 1, 1\right)$, where multiplier $\alpha$ is some constant. In this paper, we assume a default of $\alpha = \frac{1}{\sqrt{3}}$ to set the Euclidean distance between each vertex to the origin equal to 1. 

This specification ensures that each nucleotide is equidistant from every other possible nucleotide and the angle between any two nucleotides through the origin is identical.\footnote{\citet{Chang:2003} offers an alternative 3D encoding of DNA sequences, not specific to chaos games, with the same characteristics: A = $(0, 0, 1)$; T = $\left(0, \frac{2\sqrt{2}}{3}, -\frac{1}{3}\right)$; C = $\left(-\frac{\sqrt{2}}{3}, \frac{\sqrt{6}}{3}, -\frac{1}{3}\right)$; G = $\left(-\frac{\sqrt{2}}{3}, -\frac{\sqrt{6}}{3}, -\frac{1}{3}\right)$} Starting with the center at $X_0 = (0, 0, 0)$, the CGR representation $(X_n) = (x_n, y_n, z_n)$ for a DNA sequence $s_1s_2...s_n$ where $s_i \in \{A, C, G, T\}$ is built iteratively in the following manner: 
\begin{equation}
    X_n = \frac{1}{2}(X_{n-1} + W(s_n))
\end{equation}
where $W$ maps nucleotides to vertices as specified above. This algorithm is identical to that of the 2D CGR but with a new 3D system of coordinates. A 3D CGR of the mitochondrially encoded NADH:ubiquinone oxidoreductase core subunit 4 (MT-NADH) gene for the Macaca fascicular (a species of old-world monkey) is shown in Figure \ref{fig:nadh1_triangle}.

In the CGR of a DNA sequence, each nucleotide's corresponding coordinate is determined by its position in the sequence as well as by all preceding nucleotides. This encoding method preserves uniqueness and each DNA sequence will always produce a distinct CGR coordinate. However, an interesting feature of CGR is that matching substrings within two different sequences will have coordinates that begin to converge and occupy the same position as the length of the match increases; the longer the match, the closer in proximity their 3D CGR regardless of all preceding elements in the sequences. 

Take, for example, the small sequences ATCAGGCAG AND TGTAGGCAG that have mismatching nucleotides in the first 3 positions, but share the same subsequence in their remaining 6 positions. The 3D CGR of these sequences, shown in Table \ref{table:coord_example}, illustrate how although the coordinates are unique to each sequence, the matching subsequences have coordinates that are very similar and whose differences begin to disappear the longer the sequence match. This convergence in coordinates happens even when the mismatching prefixes are of unequal length. Because the trajectory of identical subsequences will converge, subsequences in the 3D CGR are, in a way, self-aligning. This makes this representation of DNA sequences amenable to alignment-based shape comparison techniques.

\begin{table}
  \fontsize{10pt}{12pt}\selectfont
  \caption{Coordinate Example}
  \centering
  \begin{minipage}{.4\linewidth}
    \centering
    \caption*{ATCAGGCAG Coordinates}
    \texttt{
      \begin{tabular}{rrr}
       $x$ & $y$ & $z$ \\
       0 & 0 & 0 \\
 -0.28868& -0.28868& -0.28868 \\ 
  0.14434&  0.14434& -0.43301\\
  0.36084& -0.21651&  0.07217\\
 -0.10825& -0.39693& -0.25259\\
 -0.34280&  0.09021&  0.16238\\
 -0.46008&  0.33378&  0.36987\\
  0.05864& -0.12178&  0.47361\\
 -0.25936& -0.34957& -0.05187\\
 -0.41835&  0.11389&  0.26274 
 \end{tabular}
    }
  \end{minipage}%
  \hfill
  \begin{minipage}{.4\linewidth}
    \centering
    \caption*{TGTAGGCAG Coordinates}
    \texttt{
      \begin{tabular}{rrr}
      $x$ & $y$ &  $z$ \\
      0 & 0 & 0 \\
   0.28868&  0.28868& -0.28868\\
  -0.14434&  0.43301&  0.14434\\
   0.21651&  0.50518& -0.21651\\
  -0.18042& -0.03608& -0.39693\\
  -0.37889&  0.27063&  0.09021\\
  -0.47812&  0.42399&  0.33378\\
   0.04962& -0.07668&  0.45557\\
  -0.26387& -0.32701& -0.06089\\
  -0.42061&  0.12517&  0.25823\\
      \end{tabular}
    }
  \end{minipage}
\hfill
  \begin{minipage}{.1\linewidth}
    \centering
    \caption*{Distance}
    \texttt{
       \begin{tabular}{c}
        \\ 
        0 \\  0.66667 \\0.50000 \\0.29167\\ 0.07292\\ 0.01823\\ 0.00456\\ 0.00114\\ 0.00028\\ 0.00007
       \end{tabular}
    }
  \end{minipage}
  \vspace{\floatsep}
  \begin{minipage}{\linewidth}
  \centering
  \includegraphics[width=.7\linewidth]{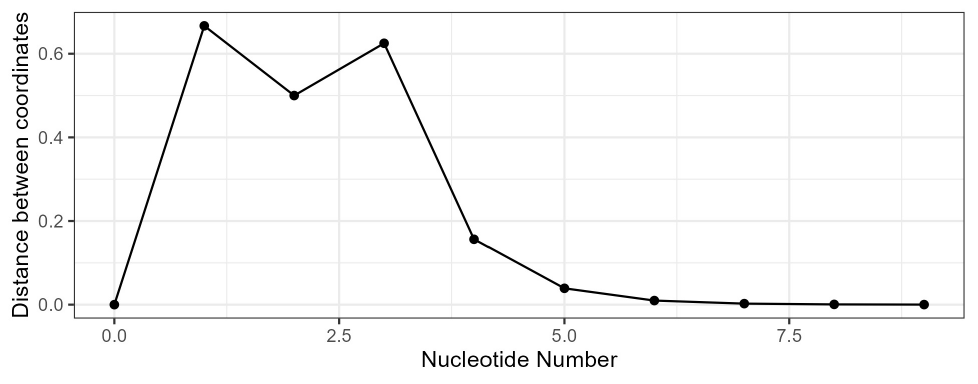}
  \\
  Plot of the distances between 3D CGR coordinates for ATCAGGCAG and TGTAGGCAG. 
  \end{minipage}
  \label{table:coord_example}
\end{table}

\subsection{Similarity and distance measures}
\label{similarity}

In order to compare DNA sequences encoded in 3D CGR, we look to distance/similarity measures that have been developed to compare shapes and volumes for inspiration. Several approaches for comparing 3D conformations, in particular with application to molecular structure, have been developed for performing shape similarity quantification.

\subsubsection{Volume Intersection}

A line of approach for comparing molecular structures in chemistry involves aligning and super-imposing structures of interest and, representing atoms as either hard or Gaussian spheres, calculating the ``distance" or ``similarity" between the two structures as a function of the integral volume of all overlapping spheres \citep{Connolly:1985, Masek:1993, Grant:1995, Grant:1996}. Such methods are intuitive, but can be computationally intensive and sub-par alignments can lead to incorrect conclusions. 

Since similar sequences will have similar 3D CGRs, we can compare pairs of sequences by comparing the shape generated by their 3D CGR trajectory. Following the spirit of the aforementioned techniques, we utilize Gaussian spheres with the Gaussian kernel density defined as:

\begin{equation}
  K{(\vv{x})} = \frac{1}{(\sigma\sqrt{2\pi})^3}e^{-\frac{||\vv{x}||^2}{2\sigma^2}}
\end{equation}
where $\sigma$, also known as the bandwidth, determines the width of the kernel and the multiplicative fraction $c = \frac{1}{(\sigma\sqrt{2\pi})^3}$ is the normalization constant. Gaussian spheres are centered at each nucleotide's 3D CGR coordinates (excluding the initial $X_0$ point) in the trajectory using the \textit{hypervolume} package in \textit{R} \citep{R:hypervolume}. The volume of the spheres are estimated using a Monte Carlo sampling approach with 5,000 sample points per nucleotide for the 3 sets of DNA sequences examined, evaluated out to four standard deviations away, with a fixed kernel bandwidth.  

The Tanimoto coefficient is used to quantify the sequence similarities: 

\begin{equation}
    T_{f, g} = \frac{O_{f, g}}{I_f + I_g - O_{f, g}}
\end{equation}
where the $I$ terms are the volumes of each entity (denoted $f$ and $g$) and the $O$ term is the overlap between the two. $T_{f, g}$ it captures the overlap in shared substructures and is sensitive to common shared features. It is generally considered to be minimally sensitive to minor variations in shape and alignment, especially when coupled with Gaussian spheres, making it appropriate for this use case. 

Since the Tanimoto coefficient is a similarity metric bounded between 0 and 1, the distance between two sequences is approximated as $T^{-}_{f, g}  = 1 - T_{f, g}$. This approach mirrors the alignment-based approaches of shape similarity metrics in  the field of cheminformatics and molecular modeling where the robustness and usefulness of the Tanimoto coefficient for comparing the shape of 3D molecules has been well-established and demonstrated in practical applications such as virtual drug screenings \citep{Taminau2008-oj, Rush2005, Hawkins2007}.

A narrower kernel bandwidth prioritizes exact matches, but can lead to very low similarity scores if it is set too low. In general, longer sequences should have smaller bandwidths to reduce chance intersections. Given that the longest sequence to be analyzed is just under 900 base-pairs long, and the distance between each vertex is only $2\sqrt{2}/\sqrt{3} \approx 1.63$, the bandwidth must be relatively narrow to produce accurate similarity metrics. The kernel bandwidth is set at 0.003 units for the results shown; for the sequences examined, this level of resolution is sufficient. However, we explore the effects of varying this parameter in the Results Section, and recommend adjusting the bandwidth value as a function of sequence lengths such that longer sequences are analyzed with smaller bandwidths to minimize chance intersections. This approach, with very narrow bandwidths, is our attempt at a point-by-point comparison of sequences in 3D CGR. 

\subsubsection{Shape signature}

An alternative method for comparing 3D configurations involves using distributional information of various characteristics to create a shape signature. In the field of chemistry, the distributional information about molecular structures, is often summarized using histograms to encode shape information. The features summarized include: 1) atomic distances (either between atoms or relative to reference points such as a molecular centroid), 2) atom triplet triangles (using information about the perimeters or angular information from surface point normal and local curvature), and 3) within-polygon distances \citep{Bemis:1992, Nilakantan:1993, Good:1995, Ballester:2007, Ballester:2011, IDSS}. These methods operate on the assumption that similar shapes will produce similar feature distributions and, therefore, shapes can be compared by comparing these constructed signatures. These methods are often fast and parallelizable. But in practice, they can result in a high number of false positives during drug discovery efforts. 

Again, taking inspiration from the field of chemistry, we also develop another similarity metric using shape signatures. The shape signatures we construct are histograms of coordinates for each axis as well as various features embedded within the 3D CGR structure. The features that we utilize are: 

\begin{enumerate}
\item Coordinates (excluding the origin) for each axis 
\item Oriented angles: 
   \begin{enumerate} 
   \item Using a 2-point sliding window of the trajectory (excluding the origin), we calculate the oriented angle formed between two vectors originating from (0, 0, 0) and ending at the points in the window.
   \item Using a 3-point sliding window of the trajectory, the first and third points describe vectors originating from the second point, which serves as the vertex, from which we calculate the angle.
   \end{enumerate}
\item Oriented distances: 
    \begin{enumerate}
    \item Using a 2-point sliding window of the trajectory (excluding the origin), oriented distances can be calculated between the two points. 
    \item Using a 3-point sliding window of the trajectory, oriented distances can be calculated between the first and third points within that window.
    \end{enumerate}
\end{enumerate}

\begin{figure}
\centering
    \includegraphics[width=.5\linewidth]{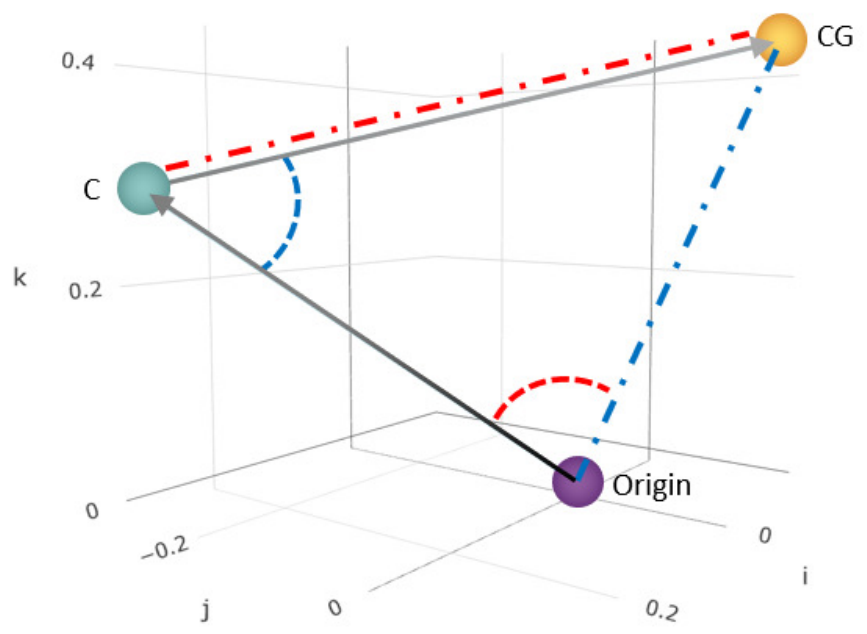}
    \caption{An illustration of the features used in the shape signature for a sequence that starts with ``CG." In addition to the coordinates of both ``C" and ``CG", the angles and edges labeled also serve as features in the shape signature. The angles in a 3-point sliding window of the trajectory is shown as the dashed curve in blue and the distance of a 3-point sliding window is shown as the dotted-dashed line in blue. The angles in a 2-point sliding window is determined using the origin as the vertex and is shown as the dashed curve in red; the corresponding distance is the dotted-dashed line in red. 
}
    \label{fig:angle_edge}
\end{figure}

Figure \ref{fig:angle_edge} shows the coordinates, angles, and distances used to generate the shape signature measurements. 

The oriented angles are calculated as: 

\begin{equation}
\theta = \mathrm{sign}(\vv{v_o} \cdot (\vv{a} \times \vv{b})) \frac{a \cdot b}{|a||b|}
\end{equation}
where the angle is between $\vv{a}$ and $\vv{b}$ and $\vv{v_o}$ is an axis of orientation. The oriented distances are calculated as:

\begin{equation}
l = \mathrm{sign}((q-p) \cdot \vv{v_o}) \sqrt{\sum_{i = 1}^3 (q_i-p_i)^2}
\end{equation}
where $p$ and $q$ are points and $p_i$ and $q_i$ are the coordinates of the two points in each dimension. As before, $\vv{v_o}$ are axes of orientation. For both the oriented angles and the oriented distances, we use (1, 0, 0), (0, 1, 0), and (0, 0, 1) as axes of orientation. 

The features are converted to a total of 15 histograms: 3 histograms hold coordinate information (with one histogram for each coordinate), 6 histograms hold oriented angle information (2 sliding-window lengths each with 3 axes of orientation), and 6 histograms to hold oriented distance information (2 sliding-window lengths each with 3 axes of orientation). Each produced histogram is centered and scaled by subtracting the mean count and dividing by the standard deviation of counts. 

To calculate the distance between each pair of sequences: 1) a Euclidean distance is calculated from each feature histogram, 2) each Euclidean distance is scaled to be between 0 and 1, and 3) the final distance is derived by pooling and averaging those distances. For the presented results, the number of bins for each histogram is set to 1/15th of the average sequence length. This parameter was selected to allow our model to be as expressive as possible without increasing the space necessary to store the histogram information than would be needed to store the original DNA sequence. For the sequences tested, the selection of this parameter did somewhat affect the overall topology of the trees, but the clustering of the leave nodes was largely invariant to this parameter. Explorations of the effects of varying bin counts is also provided in the Results section.

\subsection{Construction of phylogenetic trees}

Phylogenetic trees are built using distances derived from the methods detailed in Section \ref{similarity} to test their performance. For comparison, two baseline methods were also implemented. For the first baseline method, sequences were aligned using Clustal Omega alignment using the \textit{msa} package in \textit{R} \citep{R:msa} and distances for the aligned sequences were calculated using pairwise polymorphism p-distances as implemented by the \texttt{dist.gene()} in the \textit{ape} package \citep{R:ape}.\footnote{Evolutionary models such as the Jukes-Cantor model were not used because they are generally not amenable to deletion mutations.} For the second baseline method, distances between the sequences were calculated using k-mer distances using the \textit{kmer} package. We optimized k-mer parameters by selecting those that best reproduced the phylogenetic trees built under Clustal Omega and p-distances. We use $k = 6$ and Euclidean k-mer distances. We avoid the default distance metric, proposed by \citet{Edgar:2004}, as it produced non-monotone distances and reversals in the dendrograms, and generally resulted in poor phylogenetic reconstruction.\footnote{Other baseline approaches that were tested and excluded due to poor performance include: Nucleotide Amino Acid K-mer Vector \citep{naakv}, Extended Natural Vector method \citep{envm}, and Positional Correlation Natural Vector method \citep{pcnv}. Both the Extended Natural Vector method and the Positional Correlation Natural Vector method performed considerably worse than the presented baseline methods; the Nucleotide Amino Acid K-mer method was comparable to the k-mers method and not sufficiently different to warrant deviating away from the more general and well-known k-mers approach.}.

Phylogenetic trees are built using the neighbor-joining method as implemented by the \texttt{bionj()}\footnote{\texttt{bionj()} uses an improved implementation of the NJ algorithm by \cite{Gascuel_bionj}.} function in the \textit{ape} package \citep{R:ape}. This agglomeration method was selected because it is a popular and efficient method. Because it does not assume a constant rate of evolution across all lineages, it is applicable to a wide range of biological datasets and can be applied to the different types of DNA sequences we examine. Other agglomeration methods were tested as well but, in some instances, the reference methods performed worse than the 3D CGR methods under those alternative agglomeration approaches; to ensure that our method is robust, we always opt for modeling choices that benefit the reference methods. Ultimately, however, the choice of agglomeration method does not impact the substantive conclusions made regarding the 3D CGR method as the CGR method performed equally well across different agglomeration methods.

\subsection{Simulations of deletion and substitution mutation aggregation}

The effects of deletion mutations and substitution mutations on the 3D CGR are explored using the coding sequences (CDS) of the first exon of the $\beta$-globin gene for the chimp which consists of 105 base pairs. Point mutations are introduced to this parent sequence, increasing from 2 mutations to 76 total mutations in intervals of 2.  One child lineage receives only substitution mutations and the other receives only deletion mutations with each subsequent generation receiving two more mutations than the previous; both substitution and deletion mutation locations correspond to the exact same locations on the original parent sequence and once a location has been mutated it cannot be mutated again. Bandwidths for the volume intersection method ranged from 0.0001 to 0.0009 in intervals of 0.0001, 0.001 to 0.009 in intervals of 0.001, and 0.01 to 0.1 in intervals of 0.01. The number of bins used for histograms of the shape signature methods, at the low end, is set to equal 5\% of the average sequence length across all lineages and, at the high end, is equal the average sequence length.

Code to produce these analyses (as well as under alternative agglomeration methods) is available on \url{https://github.com/tomato-stats/3D_CGR/}.

\section{Data}

The performance of the 3D CGR signatures is evaluated using the following sequences: 

\begin{enumerate}
    \item the CDS of the first exon of $\beta$-globin gene from 11 species (listed in Table \ref{table:beta}): human, chimp, mouse, rat, gallus, gorilla, rabbit, opossum, lemur, goat, bovine, 
    \vspace{-15pt} 
    \begin{center}
    \begin{minipage}{\textwidth}
    \centering
    \captionof{table}{$\beta$-globin gene accession information}
    \setlength{\tabcolsep}{10pt} 
    \renewcommand{\arraystretch}{.85} 
    \begin{tabular}{ ll r } 
    \hline 
    Animal & ID/accession & length(bp) \\
    \hline 
    human & 455025 & 92 \\ 
    chimp & X02345 & 105 \\
    mouse & V00722 & 93 \\
    rat & X06701 & 92 \\
    gallus & V00409 & 92 \\
    gorilla & X61109 & 93 \\
    rabbit & V00882 & 92 \\
    opossum & J03643 & 92 \\
    lemur & M15734 & 92 \\
    goat & M15387 & 86 \\
    bovine & X00376 & 86 \\
    \hline\\
    \end{tabular}
    \label{table:beta}
    \end{minipage}
    \end{center}

    \item the 0.9-kb mtDNA fragments associated with the NADH subunit 4 and 5 (MT-NADH) genes from four groups of primates (listed in Table \ref{table:nadh}): four species of old world primates (Macaca fascicular, Macaca fuscata, Macaca sylvanus, Macaca mulatta), five hominoid species (Human, Chimpanzee, Gorilla, Orangutan, Hylobates), one new world primate (Saimiri Sciureus), one strepsirrhini/prosimian primate (Lemur catta), and the Tarsisus syrichta whose taxonomic classification is sometimes under the haplorrhines order (with the old world primate, new world primates, and the hominoids) and sometimes under the strepsirrhines (prosimians), 
    \vspace{-15pt} 
    \begin{center}
    \begin{minipage}{\textwidth}
    \centering
    \captionof{table}{MT-NADH gene accession information}
    \setlength{\tabcolsep}{10pt} 
    \renewcommand{\arraystretch}{.85} 
    \begin{tabular}{ ll r } 
    \hline 
    Animal & ID/accession & length(bp) \\
    \hline 
    Macaca fascicular & M22653 & 896\\
    Macaca fuscata & M2265 & 896\\
    Macaca mulatta & M22650 & 896\\
    Macaca sylvanus & M22654 & 896\\
    Saimiri sciureus & M22655S &893 \\
    Chimpanzee & V00672 &896\\
    Lemur catta & M22657 &895\\
    Gorilla & V00658 &896\\
    Hylobates & V00659 &896\\
    Orangutan & V00675O & 895\\
    Tarsisus syrichta & M22656 & 895 \\
    Human & L00016 & 896\\
    \hline \\
    \end{tabular}
    \label{table:nadh}
    \end{minipage}
    \end{center}
    
    \item a set of synthetic DNA sequences inspired by those generated by \citet{Yin:2014Fourier} (listed in Table \ref{table:syntheticseq}).\footnote{The sequences used by \citet{Yin:2014Fourier} were originally posted at https://github.com/cyinbox/GenomeDFT/blob/master/Simutation.fas. However, the sequences in the file did not match the sequences of the reported Accession IDs on the  National Center for Biotechnology Information (NCBI) and the sequence lengths on NCBI did not match the reported number of base-pairs. Additionally, the location of substitution mutations between sequences did not appear to be independently generated.} To illustrate some of the differences between these two alignment-free methods, we use this set of synthetic DNA where the true lineages of the sequences are known. The synthetic DNA sequences are constructed to examine more closely the effect of genetic recombination as well as the sensitivity of the distance metrics to parameter settings. The set of substitution mutations were generated by randomly selecting positions for mutation (without replacement) and randomly selecting nucleotides to replace the original nucleotides. The first sequence with substitution mutations contains 25 mutations; the second sequence of substitution mutations builds on the first for an additional 25 mutations, totaling 50 mutations. An analogous pair of sequences contains deletion mutations at the same positions as the substitution. Insertion mutations were performed by inserting randomly selected nucleotides into the specified positions; analogous deletion mutations were performed by deleting nucleotides in the specified positions. Transposition mutations are similar to deletion mutations, except the deleted mutations are inserted back into the original sequence in a new position.   
\end{enumerate}

    \vspace{-15pt} 
    \begin{center}
    \begin{minipage}{\textwidth}
    \centering
    \captionof{table}{$\beta$-globin gene accession information}
    \setlength{\tabcolsep}{8pt} 
    \renewcommand{\arraystretch}{.85} 
    \begin{tabular}{ ll } 
    \hline 
    Sequence name & Description \\
    \hline 
    Original & Genbank ID OR077313.1\\
    & [gene = M]\\
    & (location=[28678..29370], 693bp) \\
    $\Delta$25NT & 25 random point substitutions \\
    $\Delta$50NT & $\Delta$25NT + 25 new substitutions \\
    $-$25NT & 25 random point deletions \\
    $-$50NT & $-$25NT + 25 new deletions \\
    $-$100NT/50:149 & 100 bp deletion  \\
    \,\,\,\,\,\,& \,\,\,\,\,\,\,\,\,  from positions 50:149\\
    +100NT/50:149 & 100 bp insertion  \\
    \,\,\,\,\,\,& \,\,\,\,\,\,\,\,\,  into positions 50:149\\
    $-$100NT/50:149,$-$100NT/500:599  & Deletion of positions \\
    \,\,\,\,\,\,& \,\,\,\,\,\,\,\,\,  50:149 and 500:599 \\
    +100NT/50:149,+100NT/500:599  & 100 bp insertion in positions  \\
    \,\,\,\,\,\,  & \,\,\,\,\,\,\,\,\, 50:149 and also 500:599 \\
    $\pm$ 50NT/50:99$\rightarrow$ 150:199   & 50 bp transposition of 50:99\\
    \,\,\,\,\,\, & \,\,\,\,\,\,\,\,\, to 150:199 \\
    $\pm$ 50NT/50:99 $\rightarrow$ 350:399  & 50 bp transposition of 50:99 \\
    \,\,\,\,\,\, & \,\,\,\,\,\,\,\,\, to 350:399\\
    \hline
    \end{tabular}
    \label{table:syntheticseq}
    \end{minipage}
    \end{center}

\begin{figure}[!h]
\centering
\includegraphics[width=\textwidth]{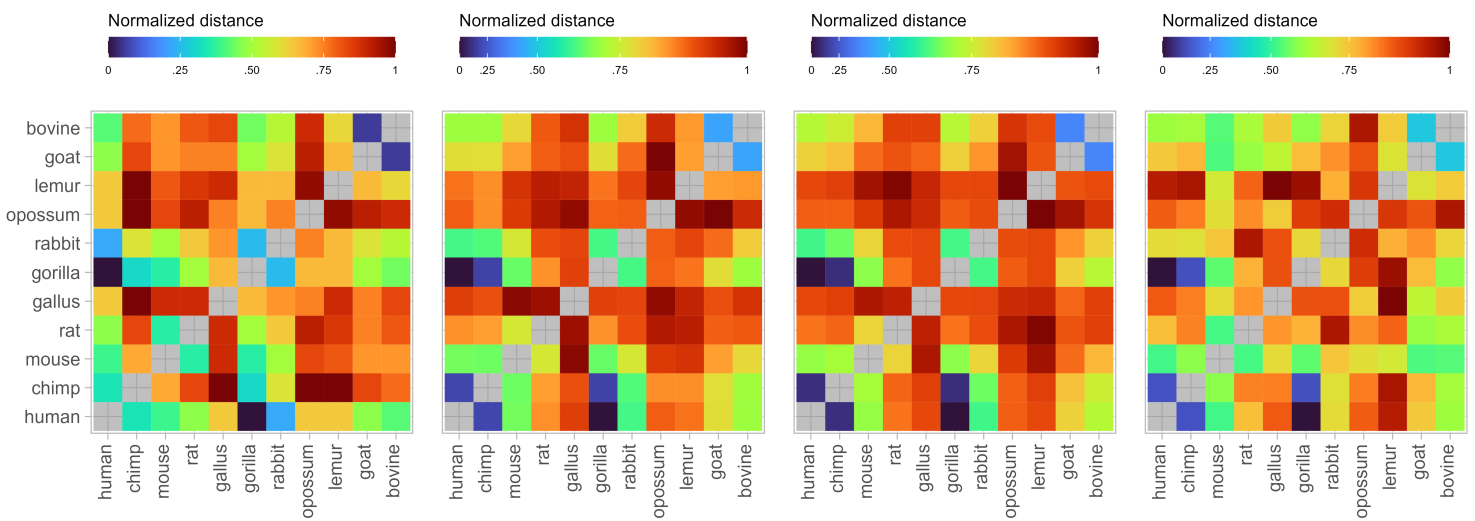}
   \begin{minipage}[t]{0.1\textwidth}
       \vspace{-10pt} 
   \end{minipage}%
   \begin{minipage}[t]{0.23\textwidth}
      \caption*{\scriptsize (a) Clustal Omega and p-distances }
      \label{subfig:beta_heatmap_a}
      \vspace{-10pt} 
   \end{minipage}%
   \begin{minipage}[t]{.25\textwidth}
      \caption*{\scriptsize (b) k-mer distances}
      \label{subfig:beta_heatmap_b}
      \vspace{-10pt} 
   \end{minipage}%
   \begin{minipage}[t]{.25\textwidth}
      \caption*{\scriptsize (c) volume intersection}
      \label{subfig:beta_heatmap_c}
      \vspace{-10pt} 
   \end{minipage}%
   \begin{minipage}[t]{0.19\textwidth}
       \caption*{\scriptsize (d) shape signature}
       \label{subfig:beta_heatmap_d}
       \vspace{-10pt} 
   \end{minipage}%
\caption{Heatmap of distances based on the CDS of the $\beta$-globin gene.}
\label{fig:beta_heatmaps}
\vspace{20pt}

\includegraphics[width=\textwidth]{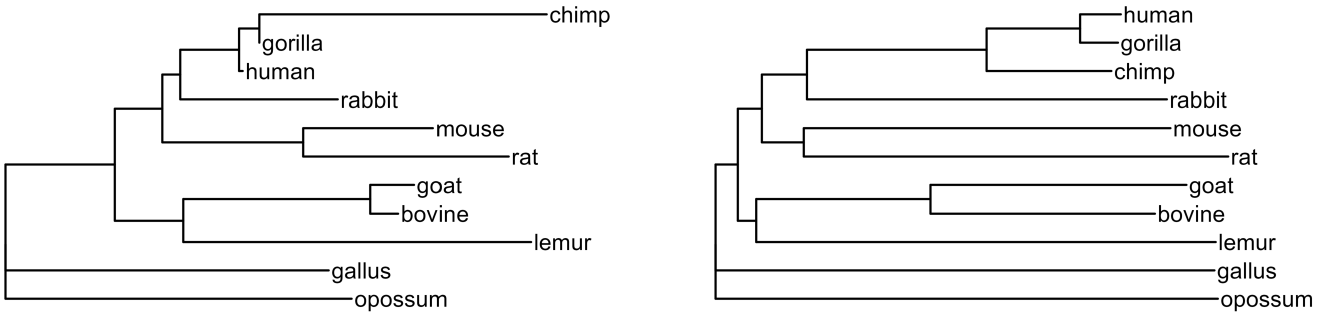}
   \begin{minipage}{0.5\textwidth}
      \caption*{\scriptsize(a) Clustal Omega and p-distances }
      \label{subfig:beta_tree_a}
      \vspace{10pt}
   \end{minipage}%
   \begin{minipage}{.5\textwidth}
      \caption*{\scriptsize(b) k-mer distances}
      \label{subfig:beta_tree_b}
      \vspace{10pt}
   \end{minipage}
\includegraphics[width=\linewidth]{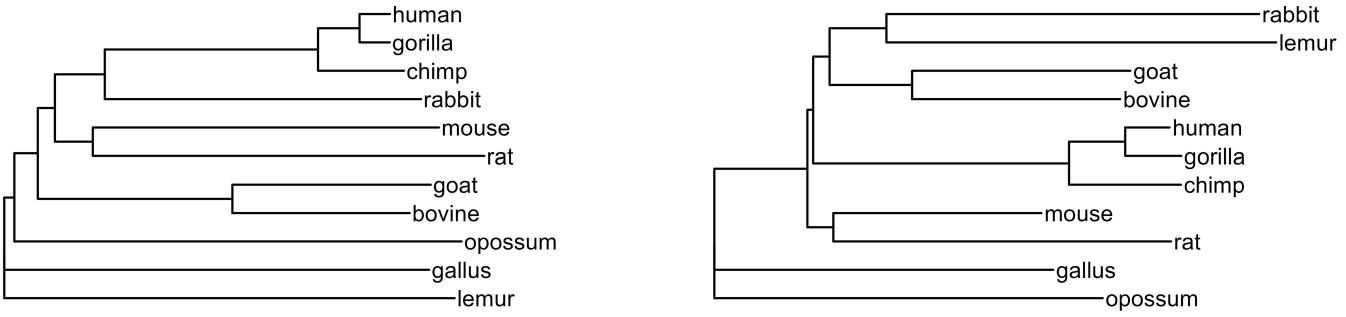}
   \begin{minipage}{.5\textwidth}
      \caption*{\scriptsize(c) volume intersection}
      \label{subfig:beta_tree_c}
   \end{minipage}%
   \begin{minipage}{0.5\textwidth}
       \caption*{\scriptsize(d) shape signature}
       \label{subfig:beta_tree_d}
   \end{minipage}%
   \hfill
\caption{Phylogenetic trees of the CDS of the $\beta$-globin gene. (a) Clustal Omega and p-distances, (b) k-mer distances, (c) volume intersection, (d) shape signature.}
\label{fig:beta_trees}
\end{figure}

For the set of these three smaller sets of sequences, we make the effort to explore specifically why our method may or may not deviate from the reference methods. In addition to these smaller datasets, we also demonstrate the shape signature approach on a substantially larger dataset---434 SARS-CoV-2 sequences obtained from NCBI's nuccore database in mid 2023 with sequences collected between 2020 and early 2023. The search term used to collect these sequences was: 
\begin{lstlisting}
txid2697049[Organism:noexp] AND collection_date=[All Fields] AND y-m-*[All Fields] AND host[All Fields] AND country[All Fields] AND isolate[All Fields] AND "complete"[All Fields] NOT partial[All Fields] AND "Sequencing Technology"[All Fields] AND ("20000"[SLEN] : "40000"[SLEN])
\end{lstlisting} 
where $y$ and $m$ are numbers to ensure collection date was included in the data pulled. Additional wrapper code prioritized wider country diversity to ensure that not all data were from the same country. Data presented are specifically from India, Japan, Brazil, and Germany as they had representation across a wide number of years and are geographically disparate. The average length of these sequences was 29,784.48 base-pairs.

\section{Results}
\label{Results}

\subsection{Coding sequences of the $\beta$-globin gene}

For the CDS of the $\beta$-globin gene, a heatmap of the distances generated using multiple sequence alignment (MSA) method Clustal Omega and p-distances, k-mer distances, intersection of volumes for Gaussian spheres in 3D CGR representation, and shape signature are provided in Figure \ref{fig:beta_heatmaps}. For all heatmaps represented, each organism's distance to itself is excluded from the plot and all remaining distances 
are scaled to be between 0 and 1; color scales are chosen to improve color diversity within each method and to better highlight the similarities and differences between methods. The resulting phylogenetic trees built from those distances are shown in Figure \ref{fig:beta_trees}. 

\begin{figure}[h]
    \includegraphics[width=\linewidth]{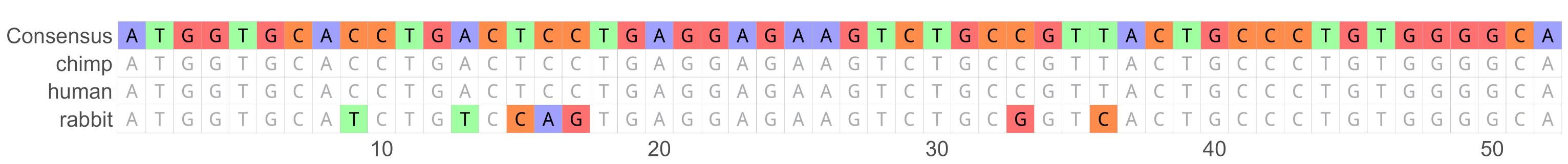}
    \includegraphics[width=\linewidth]{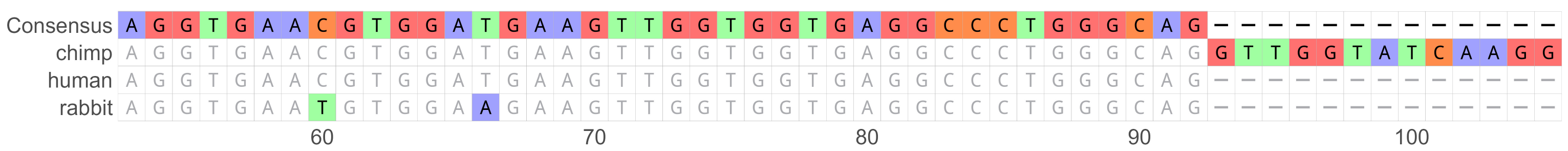}
    \caption{Sequence alignment of the CDS of the $\beta$-globin gene for the chimp, human and rabbit.}
    \label{fig:chimp_human_rabbit}
\end{figure}

The heatmaps show that the alignment-based method results in a smaller distance between the rabbit and the human than between the chimp and the human. A closer look at the rabbit, human, and chimp sequences in Figure \ref{fig:chimp_human_rabbit} reveals why this is the case: there are nine substitutions between the human and the rabbit; in contrast, the chimp sequence is identical to the human sequence with the exception of an insertion of 13 additional nucleotides. The total number of positions with mismatched nucleotides is smaller between the human and rabbit than between the human and chimp. 

In contrast, all of the alignment-free methods place the chimp closest to the human. Since the 13-nucleotide insertion at the end of the chimp sequence is likely to be a single mutation event, the total number of mutation events between the human and chimp is likely fewer than the number between the human and rabbit. In this scenario, it is not unjustified for the alignment-free methods to quantify the distance between the human and chimp as less than the distance between the human and rabbit. 

Both the Clustal Omega and k-mer reference methods produce a similar tree topology under the neighbor-joining agglomeration method despite quite disparate distance estimates. Although none of the shape-based approaches perfectly reproduce the trees built by the reference methods, they do produce near-identical leaf-node clusterings with the ungulates (bovine and goat) clustered together, the primates (human, chimp, and gorilla) clustered together, and the rodents (mouse and rat) clustered together. The lemur's total distance from all other organisms is the highest under both the volume intersection and shape signature methods, which is not concordant with the reference methods. Additionally, placement varies under the shape-based approaches; under the chosen parameter settings, the lemur clusters with no other sequence under the volume intersection method and clusters with the rabbit in the shape signature method. 

While it is technically possible to select parameter settings such that the both 3D CGR methods reproduce the general topology of the reference methods, what remains constant is that: 1) the lemur's clustering position tends to be less stable relative to other organisms and 2) the lemur is consistently amongst the set of most dissimilar organisms, regardless of parameterization, rivaled in total distance only by the oppossum (a marsupial) and the gallus (the only non-mammal organism in the set). 

The mutations observed in the lemur are uniquely different from those observed in the other organisms. Firstly, the types of mutations observed in lemurs tends to be different from those of other organisms. Existing examinations of the lemur's $\beta-$globin gene have found that the lemur lineage of the $\beta$-globin gene contains a significant excess of non-synonymous mutations over the neutral mutation rate, especially in the first 13 codons of the first exon \citep{Harris1986-no}, which we confirm. In contrast, mutations observed in other organisms are primarily synonymous mutations and/or transition mutations.

Secondly, a large proportion (9 out of 22) of the lemur's mutations are private mutations; the locations of its mutations are frequently disparate from those of other organisms and occur where all other sequences match the consensus. Excluding from consideration the chimp's insertion of 13 base-pairs at the end of its sequence, the only other organisms with private mutations are the gallus with 3 out of 22 substitution mutations being private and the opossum with only 1 out of its 23 substitution mutations being private. So when the lemur has a mutation, it is often in a location that is unusual compared to all other sequences. These singleton mutations result in the lemur being considered more distant under the shape-based methods due to the fact that its 3D CGR trajectory is distinct because it diverges at different points than the other organisms. 

The 3D CGR methods are picking up on more than just a mismatch in nucleotides; they are also implicitly capturing the location of the mismatch and identifying it as being rare. Since the lemur's mismatches tend to occur in locations that are only ever mutated in the lemur, this further separates the lemur from the rest of the organisms. A number of potential explanations have been suggested for why the lemur's mutations are so disparate from sequences such as that of the human and rabbit which have mutation levels in line with a neutral evolution model: adaptive evolution, gene conversion, and relaxed selective constraints \citep{Harris1986-no}. The 3D CGR captures the contrasting evolutionary pressures that the lemur's sequence experienced. Whether this is a desirable feature in a distance metric depends on whether a researcher is interested in assigning greater distances to rarer mutations. 

A final important note is that that both the Clustal Omega p-distances as well as the k-mer distances required an agglomeration method based on reducing total tree length such as the neighbor-joining algorithm in order to perform well. Under both the unweighted pair group method with arithmetic mean (UPGMA) and the weighted pair group method with arithmetic mean (WPGMA), the sequence-alignment approach placed the rabbit between the chimp and the human/gorilla cluster. Under both UPGMA and WPGMA, the k-mer distance approach consistently resulted in the goat/bovine cluster being placed between the rat and the mouse; increasing the k-mer length up to 10 did not resolve the issue. None of these issues were observed with either of the shape signature methods whose leaf-node clusterings remained robust and consistent with expectations. This is noteworthy because the k-mer distances look very similar to the volume intersection distances. Yet the subtle differences between the two are sufficient to ensure that the volume intersection tends to perform better. This is indicative that the volume intersection approach is capturing additional information that is not captured by the k-mers approach. 

\subsection{MT-NADH gene}

\begin{figure}[!ht]
\centering
\includegraphics[width=\textwidth]{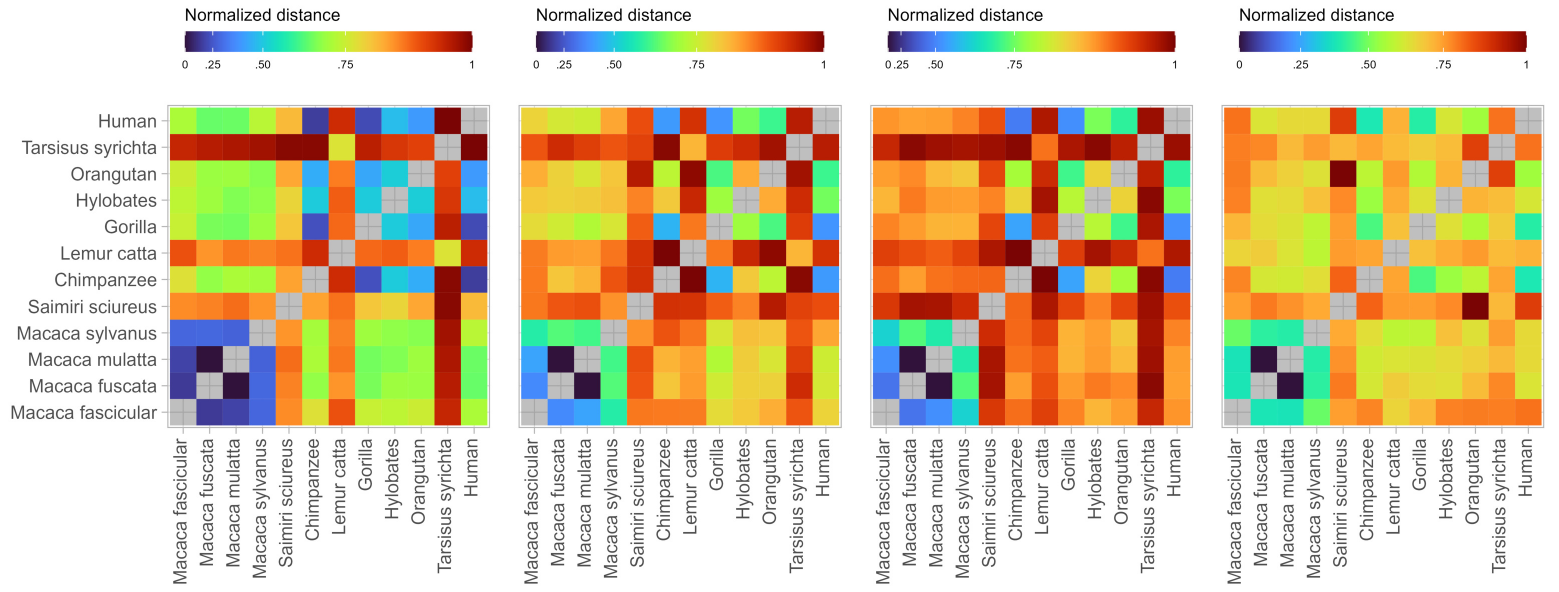}
   \begin{minipage}[t]{0.22\textwidth}
      \caption*{\scriptsize  (a) Clustal Omega and p-distances }
      \label{subfig:nadh_heatmap_a}
      \vspace{-10pt} 
   \end{minipage}%
   \begin{minipage}[t]{.23\textwidth}
      \caption*{\scriptsize (b) k-mer distances}
      \label{subfig:nadh_heatmap_b}
      \vspace{-10pt} 
   \end{minipage}%
   \begin{minipage}[t]{.25\textwidth}
      \caption*{\scriptsize (c) volume intersection}
      \label{subfig:nadh_heatmap_c}
      \vspace{-10pt} 
   \end{minipage}%
   \begin{minipage}[t]{0.19\textwidth}
       \caption*{\scriptsize (d) shape signature}
       \label{subfig:nadh_heatmap_d}
       \vspace{-10pt} 
   \end{minipage}%
\hspace{-20pt}
\caption{Heatmap of distances based on the MT-NADH gene.}
\label{fig:nadh_heatmap}
\end{figure}

\begin{figure}[!ht] 
\centering
\includegraphics[width=\linewidth]{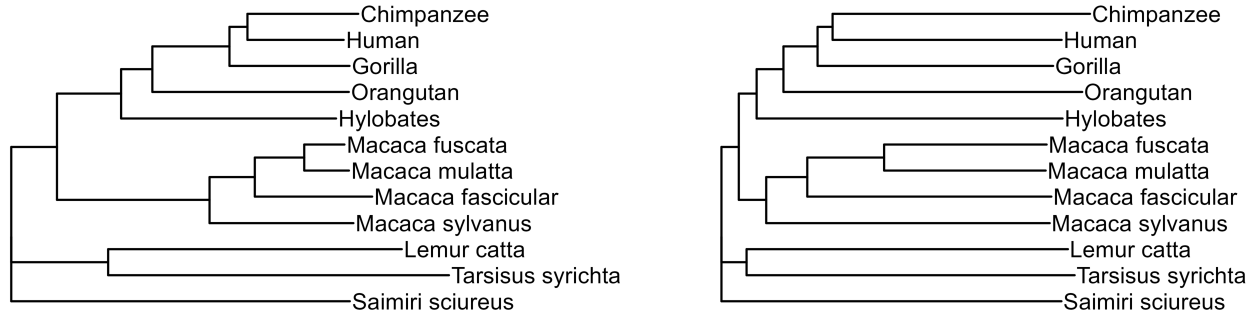}
   \begin{minipage}{0.5\textwidth}
      \centering
      \caption*{\scriptsize (a) Clustal Omega and p-distances }
      \label{subfig:nadh_tree_a}
   \end{minipage}%
   \begin{minipage}{.5\textwidth}
      \centering
      \caption*{\scriptsize (b) k-mer distances}
      \label{subfig:nadh_tree_b}
   \end{minipage}
\includegraphics[width=\linewidth]{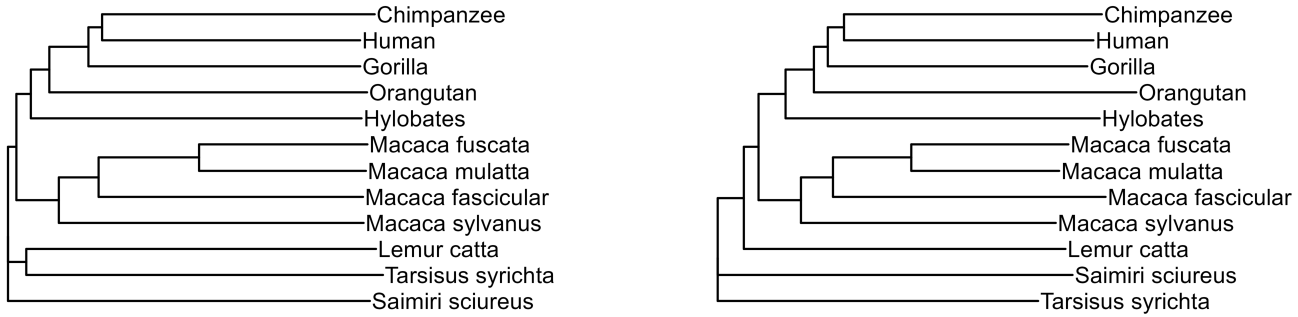}
   \begin{minipage}{.5\textwidth}
      \centering
      \caption*{\scriptsize (c) volume intersection}
      \label{subfig:nadh_tree_c}
      \vspace{-10pt} 
   \end{minipage}%
   \begin{minipage}{0.5\textwidth}
       \centering
       \caption*{\scriptsize (d) shape signature}
       \label{subfig:nadh_tree_d}
       \vspace{-10pt} 
   \end{minipage}
\caption{Phylogenetic trees of the CDS of the MT-NADH gene.}
\label{fig:nadh_trees}
\end{figure}

For the MT-NADH gene, a heatmap of the distances generated using MSA with Clustal Omega and p-distances, k-mer distances, intersection of volumes for Gaussian spheres in 3D CGR representation, and shape signature are provided in Figure \ref{fig:nadh_heatmap}. The resulting phylogenetic trees built from those distances are shown in Figure \ref{fig:nadh_trees}. 

The distributions of distances are quite varied amongst the different approaches. But although the magnitude of the distances between the methods can be quite different, the patterns of distances are somewhat similar across the methods. For example, the darker red blocks for most pairwise comparisons with Tarsisus syrchita, Lemur catta, and Saimiri sciureus indicate that these primates tend to be the most different from all others regardless of method chosen. And the blue block of old world primates also appears fairly consistently across the different methods. In general, the alignment-free methods tend to consider the old world primates to be much more distant from the hominoids than the alignment-based method. For the k-mers approach, this is likely due to the fact that multiple single-nucleotide substitutions have much larger downstream consequences for k-mer counts. For the volume intersection approach, the disparity is similar as multiple single-nucleotide substitutions in close, but not adjacent positions, can repeatedly alter the 3D CGR trajectory that, when paired with a very small bandwidth, can minimize the calculated volume overlap/similarity. 

All methods depicted in Figure \ref{fig:nadh_trees} correctly cluster the old world primates together and the five hominoid species together. Both the k-mer distance and the distances derived from the volume intersection method produce a tree topology that closely resembles the one built under alignment and p-distances. The shape signature distances also produce very similar trees with the exception of the placement of the Lemur Catta as, unlike the other methods, it is not clustered with the Tarsisus syrichta. However, as there remains some differences of opinion regarding whether the tarsiers truly belong under the group of strepsirrhines/prosimians (tarsiers do not have moist noses while all other strepsirrhines/prosimians do), the difference in the trees appear to reflect ongoing discussions regarding the correct taxonomy of primates. The taxonomy of the primates is likely to be revised in subsequent years as more DNA data is collected and sequenced and so the topology under the shape signature may yet still turn out to accurately reflect biological evolution. 

\subsection{Synthetic DNA sequence}

\begin{figure}[!ht] 
\centering
\includegraphics[width=\textwidth]{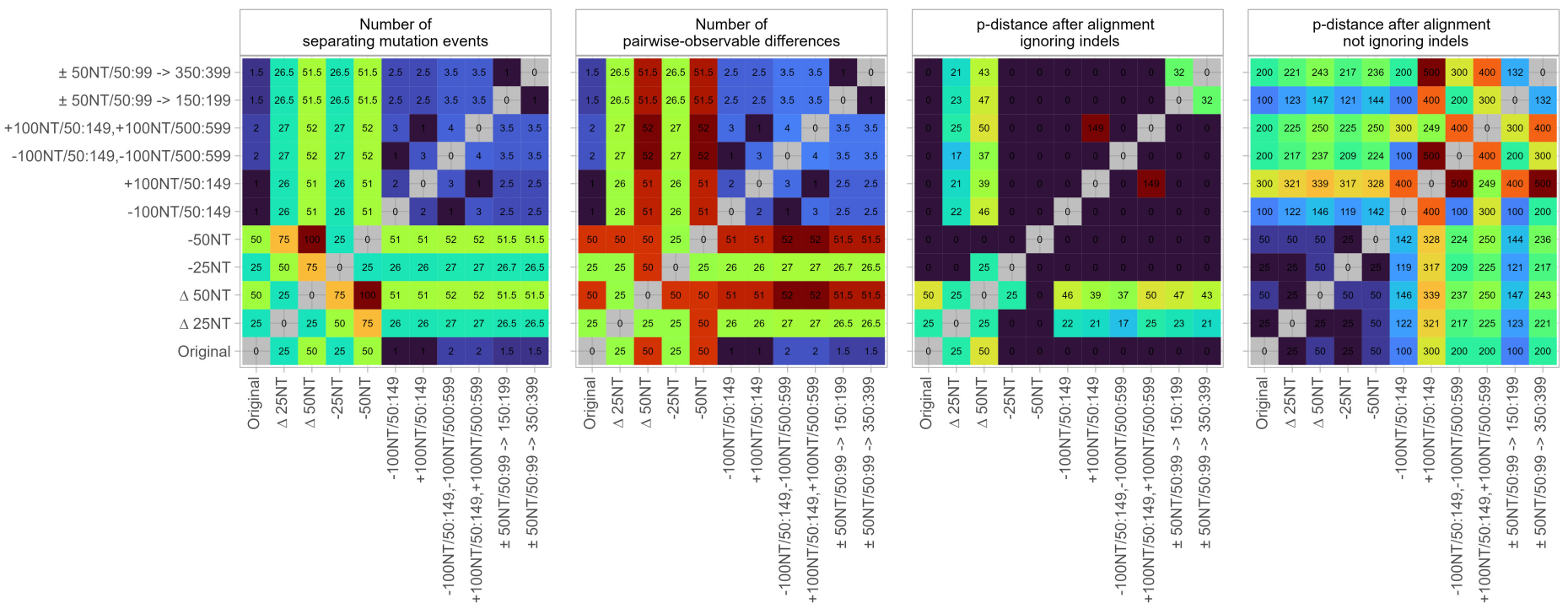}
\caption{Plots of distances based on ground-truth information: 1) the number of mutation events between each pair of sequences; 2) the upper-bound of pairwise-observable differences between any pair of sequences under ground-truth alignment; 3) the p-distance/count of mismatching nucleotides under ground-truth alignment without penalty for gap insertions; 4) the p-distance/count of mismatching nucleotides under ground-truth alignment with penalty for gap insertions.}
\label{fig:ground_truth}
\end{figure}

For the synthetic DNA sequences, we provide some ground-truth information in Figure \ref{fig:ground_truth}. In the first panel of Figure  \ref{fig:ground_truth}, we provide the exact number of separating mutation events based on how the sequences were constructed. For instance, the number of mutation events separating $\Delta$50NT from $+$100NT/50:149, a sequence with a 100-contiguous nucleotide insertions, is exactly 51; one mutation to return $+$100NT/50:149 to the original parent sequences and 50 mutations to $\Delta$50NT. The number of mutation events separating the $\Delta$50NT from the sequences with transpositions (e.g., $\pm$ 50NT/50:99 $\rightarrow$ 150:199) is 51 if the transposition is counted as a single mutation and 52 if the deletion and insertion events are each counted separately; for simplicity, we counted a transposition as 1.5 mutation events. The sequences with the greatest number of separating events are $\Delta$50NT and $-$50NT; both descend from the parent sequences with 50 mutations each and, thus, there are 100 total mutation events separating the two from one another. 

In the second panel of Figure \ref{fig:ground_truth}, we provide the theoretical number of mutation events observable between any pair of sequences under ground-truth alignment. When examining any pair of sequences without contextual knowledge of the parent sequences or information about how those sequences came about, the number of observable differences must be less than or equal to the number of mutation events. For instance, by construction, there are exactly 100 mutation events separating $\Delta$50NT and $-$50NT (50 mutation events for each sequences to return to the original parent sequence). However, in the absence of any context or information about the parent sequence and, given that the location of deletions/substitutions are identical, it would appear that there are only 50 mutation events separating those sequences when those two sequences are correctly aligned. For concreteness, consider the toy example where we compare the sequences ACG and A-G (where the hyphen denotes a deletion mutation), both mutated from their shared ancestor ATG. These two sequences are separated by exactly two mutation events--one substitution event and one deletion event. However, without information about the shared ancestor, we only observe a single difference between ACG and A-G at the second nucleotide position. For comparisons between sequences such as $\Delta$50NT and one of the sequences with 100-contiguous nucleotide deletions, the number of observable differences is 51 if none of the substitution mutations occur within the deleted region and fewer than 51 otherwise. This second heatmap essentially serves as an upper-bound on the number of observable mutations under correct pairwise alignment. 

In the third and fourth panel of Figure \ref{fig:ground_truth}, we perform MSA of all sequences manually, using knowledge of each sequence's construction. Due to disparate sequence lengths, multiple gap insertions are necessary to align sequences. The original sequences ranged from a minimum of 693 base-pairs to a maximum of 893 base-pairs. The aligned sequences resulted in all sequences being 993 base-pairs long. The third panel heatmap shows the number of mismatching nucleotides after alignment, but without penalizing mismatches due to gaps in the sequences. The fourth panel heatmap shows the same, but with penalties for gaps in the sequences. The fourth panel corresponds to p-distances under a perfect MSA method. 

\begin{figure}[!h] 
\centering
\includegraphics[width=\linewidth]{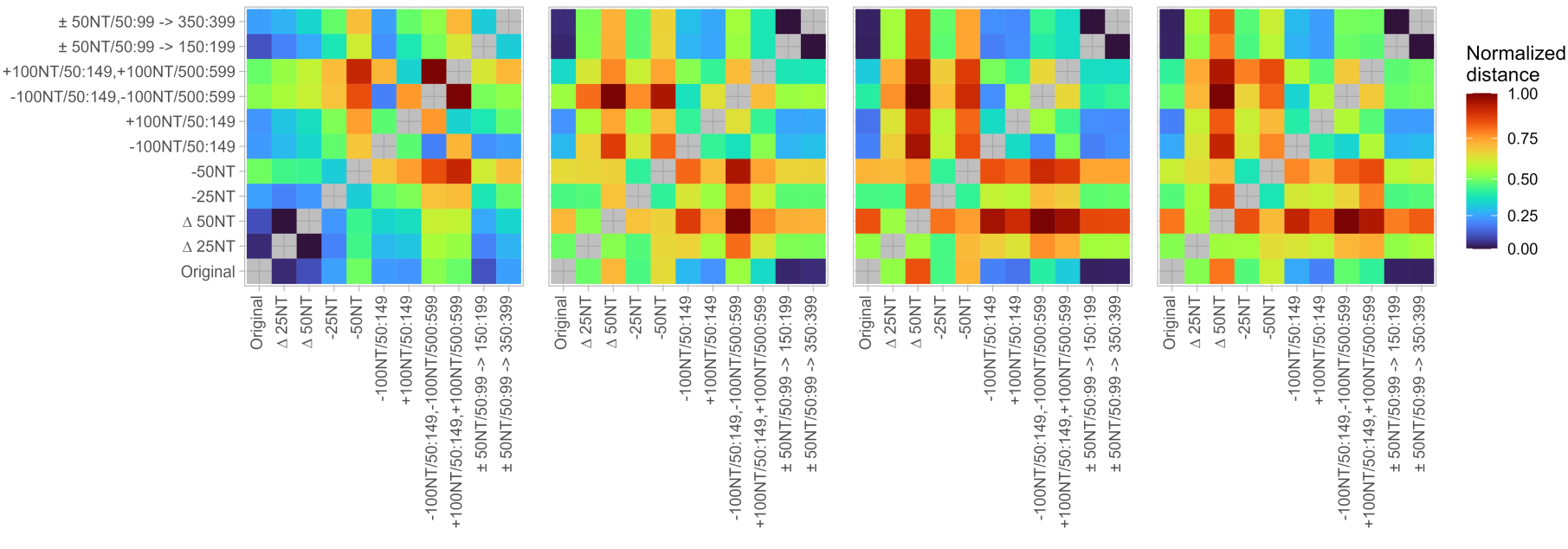}
   \begin{minipage}{0.14\textwidth}
       \vspace{-10pt} 
   \end{minipage}%
   \begin{minipage}[t]{0.25\textwidth}
      \caption*{\scriptsize (a) Clustal Omega and p-distances }
      \label{subfig:syn_heatmap_a}
      \vspace{-10pt} 
   \end{minipage}%
   \begin{minipage}[t]{.21\textwidth}
      \caption*{\scriptsize (b) k-mer distances}
      \label{subfig:syn_heatmap_b}
      \vspace{-10pt} 
   \end{minipage}%
   \begin{minipage}[t]{.22\textwidth}
      \caption*{\scriptsize (c) volume intersection}
      \label{subfig:syn_heatmap_c}
      \vspace{-10pt} 
   \end{minipage}%
   \begin{minipage}[t]{0.19\textwidth}
       \caption*{\scriptsize (d) shape signature}
       \label{subfig:syn_heatmap_d}
       \vspace{-10pt} 
   \end{minipage}%
\caption{Heatmap of distances based on the synthetic DNA sequences.}
\label{fig:syn_heatmap}
\end{figure}

\begin{figure}[!h] 
\centering
\includegraphics[width=\linewidth]{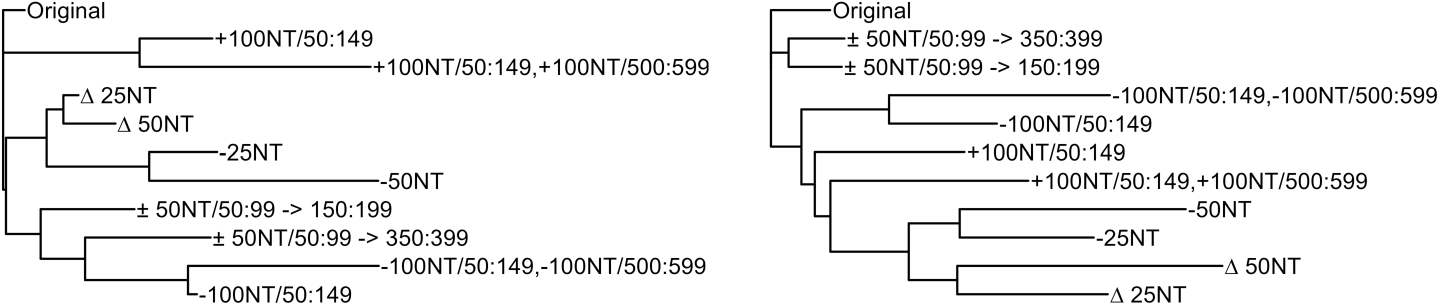}
\vspace{10pt}
   \begin{minipage}{0.5\textwidth}
      \caption*{\scriptsize (a) Clustal Omega and p-distances }
      \label{subfig:syn_tree_a}
      \vspace{-10pt} 
   \end{minipage}%
   \begin{minipage}{.5\textwidth}
      \caption*{\scriptsize (b) k-mer distances}
      \label{subfig:syn_tree_b}
      \vspace{-10pt} 
   \end{minipage}
\includegraphics[width=\linewidth]{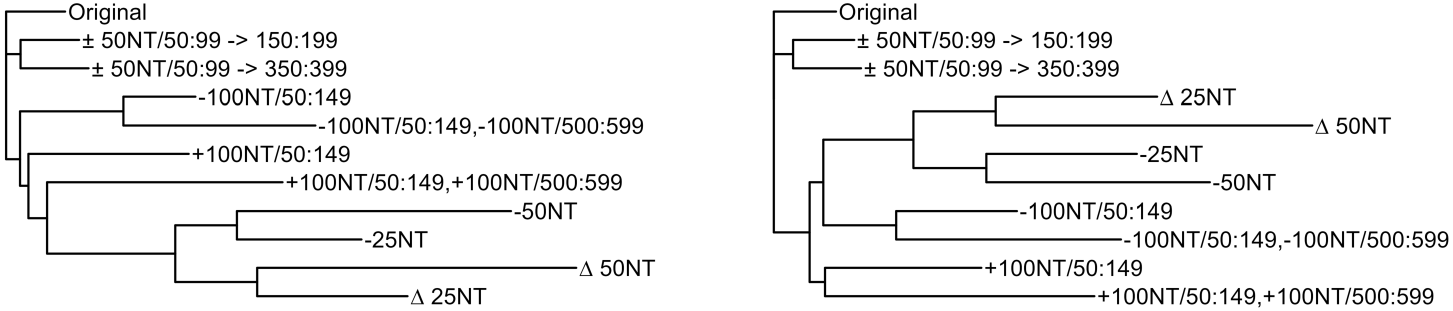}
   \begin{minipage}{.5\textwidth}
      \caption*{\scriptsize (c) volume intersection}
      \label{subfig:syn_tree_c}
      \vspace{-10pt} 
   \end{minipage}%
   \begin{minipage}{0.5\textwidth}
       \caption*{\scriptsize (d) shape signature}
       \label{subfig:syn_tree_d}
       \vspace{-10pt} 
   \end{minipage}
\caption{Phylogenetic trees of the synthetic DNA sequence.}
\label{fig:syn_trees}
\end{figure}

For comparison to the ground-truth data of the synthetic sequences, a heatmap of the distances generated using Clustal Omega with p-distances, k-mers, intersection of volumes for Gaussian spheres in 3D CGR representation, and shape signature are provided in Figure \ref{fig:syn_heatmap}. Though not especially meaningful, the resulting phylogenetic trees built from those distances are shown in Figure \ref{fig:syn_trees} for completeness. 

From comparing the ground-truth plots and the distance plots, we can see that the Clustal Omega method with p-distances produces distances that are similar to the fourth ground-truth plot, p-distance after alignment (not ignoring indels). The shape-based methods produce distances similar to the heatmap for the number of pairwise-observable difference. The k-mers method produces distances that are also similar to the number of pairwise-observable differences, but the resemblance is weaker than for the shape-based methods. 

For all distance metrics, as the number of random point substitutions and deletions increases, the distance from the original parent sequence also increases. However, under p-distances with sequence alignment, the distance between the parent sequence and the sequence with 25 random nucleotide substitutions is much smaller than the distance generated from deletion of those same 25 nucleotides even though, in theory, they should be identical. This disparity is due to the erroneous alignment for sequences with deletion mutations. The misalignment is so influential that the sequence with 50-nucleotide substitutions is considered closer to the parent sequence than the sequence with only 25 random deletions. In contrast, all alignment-free methods produce distances where the 25/50-nucleotide substitution is much more commensurate with its deletion counterpart. 

Another difference between the alignment-based method and the alignment-free methods is the effect of larger contiguous mutations on the estimated distances. For alignment-free methods, the distances from the parent sequence generated by 100-nucleotide indel mutations are smaller than those generated by the 25- and 50-nucleotide random substitutions and deletions; the opposite is the case under the alignment-based approaches. As in the case of the CDS of $\beta$-globin gene for the chimp, these large contiguous mutations were single mutation events and so the smaller distance reflects the history of mutation events. 

Since the differences between the k-mer distances and the volume intersection method were so subtle for the real DNA sequences examined, we take advantage of these synthetic sequences with its larger recombination mutations to more closely examine the commonalities between the two methods as well as where they diverge. For all alignment-free methods, sequences that have transposition mutations remain closest to the original parent sequence as all of the original elements of the parent sequence are still retained in the child after transposition; additionally, the two sequences with transposition mutations produce very similar distances, which can be seen in the first two rows of the heatmaps. In contrast, under the alignment-based method, the distance generated by the transposition sequence $\pm$ 50NT/50:99 $\rightarrow$ 350:399 is inadvertently considered to be more distant than the sequence $\pm$ 50NT/50:99 $\rightarrow$ 150:199; this is counter-intuitive because the two mutations are quite similar and affect the same number of nucleotides. 

The differences between the shape-based methods and the k-mer distance are much more subtle, but still noteworthy. For example, under k-mer distances, the sequence that was generated by two separate deletion events ($-$100NT/50:149,$-$100NT/500:599) produces a much greater distance from the parent sequence than the sequence generated by two similar insertion events ($-$100NT/50:149,$-$100NT/500:599). However, the total number of mutation events is identical between the two sequences; the disparity in the k-mer distance is due merely to the changing distribution of the k-mers resulting from the recombination of sequences. In contrast, these two sequences produce comparable distances to each other for the shape-based methods; their distance from the original parent sequence is also very small, better reflecting the low number of mutation events. 

Another noticeable difference between the shape-based distances and k-mer distances is in, for instance, all the pairwise distances between the sequence with 50 random point substitutions ($\Delta$50NT) and all other sequences. As stated earlier, the number of modifications between $\Delta$50NT and any sequences with large contiguous deletions, insertions, and transpositions is always between 51 and 52; the same is true for the $-$50NT sequence. The shape-based approach accurately identifies the two sequences, $\Delta$50NT and $-$50NT, as the most consistently different in all pairwise comparisons with other sequences (excluding their progenitor sequences, $\Delta$25NT and $-$25NT), as shown by the red rows and columns corresponding to $\Delta$50NT and $-$50NT. But, additionally, the range of observed distances between them and the sequences with large recombinant mutations is also small, reflecting the narrow range for the number of separating mutation events. The shape-based distances for $\Delta$50NT (and $-$50NT) to all other sequences with recombinant mutations are consistently high in magnitude when compared against the other sequences. In contrast, for the k-mer method with $k = 10$, the distance for $\Delta$50NT (and $-$50NT) is largest when it is evaluated against the sequence with two large deletion events, $-$100NT/50:149,$-$100NT/500:599, but does not stay similarly high for the other recombinant mutation sequences. Overall, we can see that the color patterns in the shape-based methods closely follow the heatmap for the number of pairwise-observable differences of Figure \ref{fig:ground_truth}, which is not seen in the k-mer method. 

The resulting phylogenetic trees for these synthetic sequences reflect not only the distances shown in the heatmap, but the effectiveness of the neighbor-joining agglomeration method. For all trees, the pair of sequences with overlapping substitution mutations are clustered together and those with overlapping deletion mutations are always clustered together. The sequences with shred insertion mutations are also quite close in the tree. As in the case of the $\beta$-globin gene, the agglomeration method used on the synthetic DNA does affect the sequence alignment tree; under UPGMA, the sequence alignment approach fails to place the 50-nucleotide deletion sequence near the 25-nucleotide deletion sequence, its progenitor/antecedent sequence. The other methods are less affected by agglomeration method. 

\begin{figure}[h] 
\centering
\includegraphics[width=\linewidth]{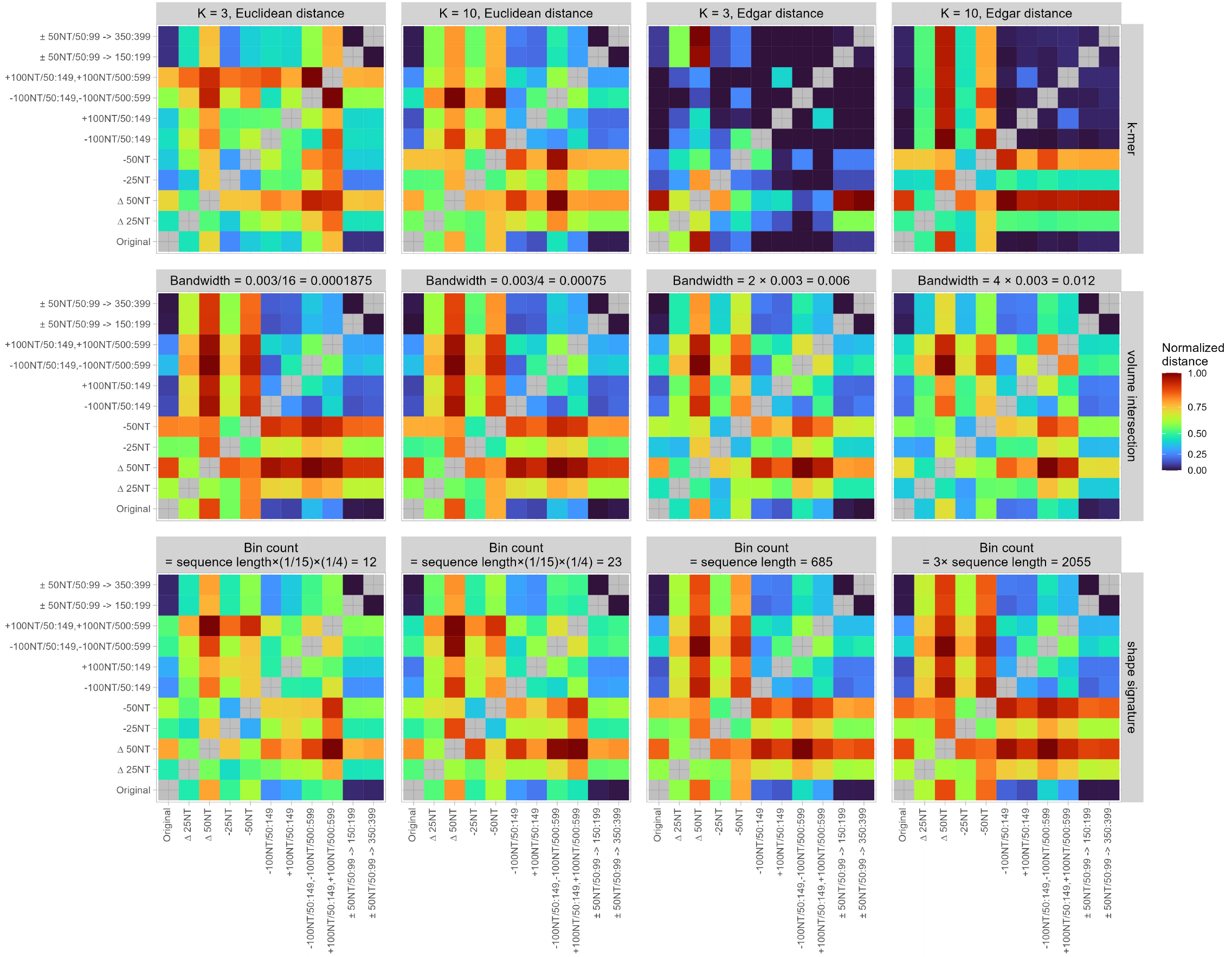}
\caption{Heatmap of distances for varying parameter values associated with k-mer distances (top row), volume intersection (middle row), and shape signature (bottom row). Though similar, the subtle differences are quite important and are indicative of the shape-based approaches ability to better capture the number of separating mutation events over the k-mer distances, which is further detailed in the text.}
\label{fig:varying-params}
\end{figure}

In Figure \ref{fig:varying-params}, we show heatmaps for the distances of the alignment-free approaches under alternative parameter values. Figure \ref{fig:varying-params} shows heatmaps for: 1) k-mers of length 3 and 10 and with distances set to either Euclidean or the default \cite{Edgar:2004} distance,\footnote{The k-mers with $k>10$ were computationally intractable.} 2) volume intersection distances with bandwidths equal to $0.003\left(\frac{1}{16}\right), 0.003\left(\frac{1}{4}\right), 2\times 0.003$, and $4\times 0.003)$; and 3) shape signature distances for varying histogram bin counts as a fraction, $\frac{1}{60}$th and $\frac{1}{30}$th, or a multiple, 1 and 2, of the average sequence length (equivalent to changing the number of bins in each feature from the original 46 to 12, 23, 685, and 1370). Much more granular parameterizations were explored; the subset presented here are sufficient to illustrate the observed changes in distances as a function of parameter settings. 

Most notably, the distances and phylogenies derived from k-mers is quite different under the Edgar method. As expected, there is a vast deterioration in the quality of the distances when reducing k-mer length down to 3. And there is an improvement in the quality of the distances when increasing the k-mer length to 10 as the pairwise distances between recombined sequences and $\Delta$50NT increase and they become relatively more equal; the same applies to the pairwise distances between recombined sequences and $-$50NT. But the noted elevation in distance between the 50-nucleotide substitution and the 100-nucleotide insertion into positions 50:149 and 500:599 remains unchanged. Additionally, there is a limit to how much the distances can be improved under k-mers given that longer k-mer lengths become computationally difficult as the number of unique k-mers is $4^k$. 

Both the volume intersection method and the shape signature method rely only on one parameter. Both decreasing the size of the bandwidth as well as increasing the number of histogram bins improve the distance metrics by bringing their relative distances more in line with our expectations shown in the ground-truth plots. However, in contrast to the k-mers approach, it is not more computationally demanding to change the bandwidth of the volume intersection method and the computational demands of the shape signature is driven primarily by the sequence length rather than bins (though practical issues like memory usage and cache performance can cause slowdowns for large bin counts). The changes in the heatmaps as a function of parameter values are subtle and gradual, but when we increase the granularity of the shape parameters the heatmaps begin to better reflect the true number of observable mutation events separating the sequences. 

Finally, not only are the shape-based methods cheap to reparameterize, they produce relative distances that are quite consistent across different parameterizations. We show extremely wide parameter settings in order to demonstrate the effects of parameter selection; but we can see that the heatmaps for our presented results (Figure \ref{fig:syn_trees}), with a bandwidth of 0.003 for the volume intersection method and bin count equal to 1/15th the average sequence length (46 bins), are quite similar to heatmaps of parameters with the much higher resolution in Figure \ref{fig:varying-params}. When testing different parameter settings, we observed that the heatmaps were relatively stable unless very extreme values were chosen. This also translates into much more stable phylogenies that are largely invariant, especially at leaf node clusterings, to parameter specifications. Both of the baseline methods using sequence alignment and k-mers required particular agglomeration methods in order to perform well across all 3 DNA sequences examined here; this was not the case for the shape-based methods. 

\subsection{The effects of mutation type on 3D CGR similarity metrics}

\begin{figure}[ht] 
\centering
\includegraphics[width=\linewidth]{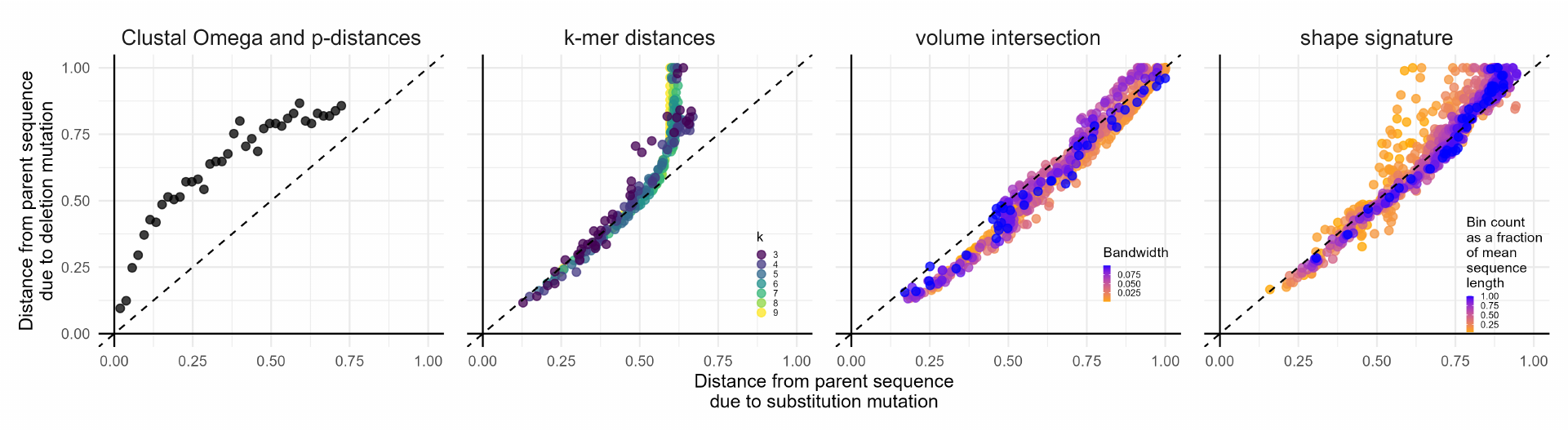}
\caption{The effects of mutation aggregation on distances. The x-axis shows the distance induced by substitution mutations and the y-axis shows the distance induced by deletion mutations at the same locations. The distances have been scaled to be between 0 and 1 to facilitate comparisons: the distance for Clustal Omega and p-distance is scaled by dividing the p-distances by the sequence length of the original parent sequence; the distance based on k-mers is scaled by dividing by the largest observed distance value within each $k$; the distance for the volume intersection method is obtained by dividing by the largest observed distance given a fixed bandwidth; and the distance for the shape signature method is obtained by dividing by the largest observed distance given a fixed bin count.}
\label{fig:mutation_agg}
\end{figure}

The overall summary of results from mutation aggregation simulations for both deletion and substitution mutations are plotted in Figure \ref{fig:mutation_agg}. The figures show that, on average, the distance calculated by shape-based methods for a deletion mutation is roughly mirrored by the distance created by a substitution mutation. In contrast, the method based on sequence alignment heavily relies on correct sequence alignment; since a deletion mutation creates greater difficulty in alignment, the effect is that a deletion mutation often appears to induce a greater distance than an equivalent substitution mutation. A relationship between k-mer distance due to substitution versus deletion mutations is similar to shape-based methods for low k-mer lengths. The bulk of the linear relationship observed between substitution and deletion mutations occur for the lowest values of $k$. As the value of $k$ increases, the shorter the linear portion of the relationship becomes. Increasing the $k$ comes at the cost of the linear relationship as deletion mutations become increasingly more penalized for higher values of $k$. However, keeping $k$ low to preserve the linearity leads to poorer characterization of the DNA sequences. For k-mers, as mutations accumulate, the distance generated by a deletion mutation deviates away from that generated by a substitution mutation regardless of the length of k; it merely takes more mutations for lower $k$ values to do so. The k-mer distances calculated under the Edgar method (not plotted) does produce a linear relationship between substitution and deletion mutation distances, but, as noted previously, the Edgar distances also perform poorly for phylogenetic analyses. 

\begin{figure}[ht] 
\includegraphics[width=\linewidth]{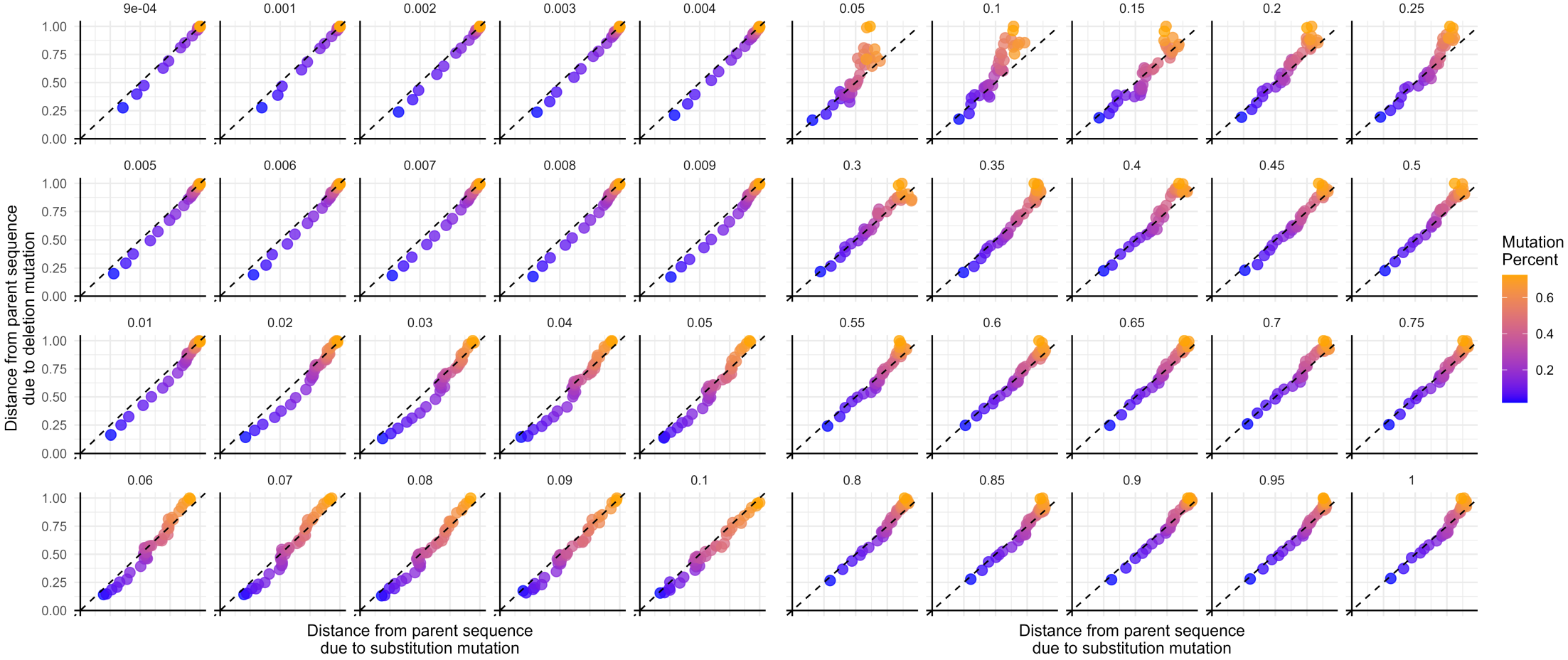}
   \begin{minipage}{.45\textwidth}
      \caption*{(a) volume intersection}
      \label{subfig:mut_agg_vi_details}
      \vspace{-10pt} 
   \end{minipage}%
   \begin{minipage}{.55\textwidth}
       \caption*{(b) shape signature}
       \label{subfig:mut_agg_ss_details}
       \vspace{-10pt} 
   \end{minipage}%
\caption{The effects of mutation aggregation on distances as a function of shape parameters. The x-axis shows the distance induced by substitution mutations and the y-axis shows the distance induced by deletion mutations at the same locations. The distances have been scaled to be between 0 and 1 to facilitate comparisons. Each facet for the volume intersection method shows the effect of bandwidth value (smaller bandwidths are not shown as they are not qualitatively different) and each facet for the shape signature method shows the effect of the number of bins used per histogram (as a fraction of average sequence length). }
\label{fig:mutation_agg-shape}
\end{figure}

A more detailed breakdown of the shape-based methods are provided in Figure \ref{fig:mutation_agg-shape} which shows how the shape parameters (bandwidth for the volume intersection method and the number of bins for the shape signature method) affect the induced distances. Results from the volume intersection method are relatively straightforward; the distance created by a deletion mutation is roughly mirrored by the distance created by a substitution mutation, especially when bandwidths are small. The choice of bandwidth affects how quickly mutations approach a distance of 1 between a parent and child sequence; smaller bandwidths, which reduce the size of the Gaussian spheres and reduce the likelihood of intersections, approach a distance of 1 more quickly than larger bandwidths. For the shape signatures, a similar dynamic is observed as the number of histogram bins increase. Although the CGR for each sequence is unique, the shape signature is not necessarily unique; however, increasing the number of bins used in the histogram reduces the chance that two different sequences have the same shape signature. Both smaller bandwidths and greater number of bins correspond to capturing more detail in the data and so smaller bandwidths and greater bin counts both correspond to greater congruence between the distance created by a deletion mutation and that created by a substitution mutation. However, the shape signature method can potentially suffer when the bin counts are increased so much that very similar CGR values fall into separate bins and cannot contribute to the estimated similarity. The volume intersection method with its Gaussian spheres is better at allowing for small misalignments due to the lack of a hard edge cutoff. 

The relationship between distances caused by deletions and distances caused by substitutions only start to deviate for the shape signature method when: 1) a large proportion of the original sequence has been mutated and 2) the shape parameter chosen is not very fine-grained (i.e., when bin counts are low). Additionally, larger bandwidths and fewer bin counts also cause greater variability in estimated distances due to deletion and substitution.

\section{Application}

\begin{figure}
\includegraphics[width=\linewidth]{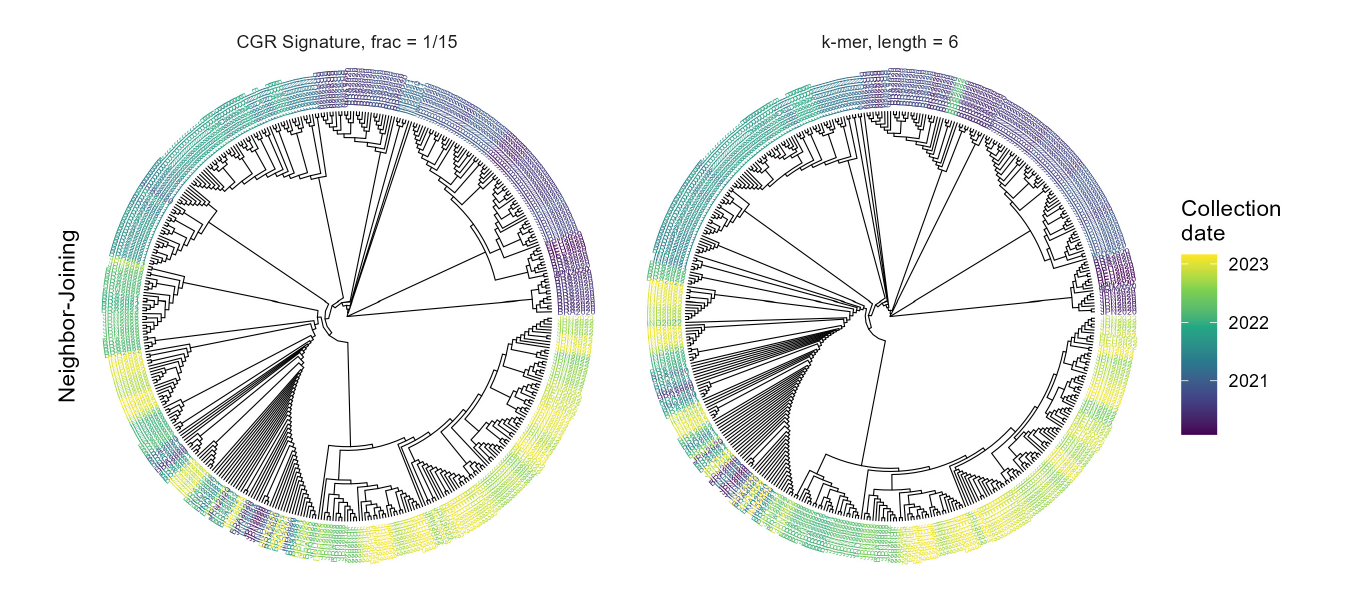}
\includegraphics[width=\linewidth]{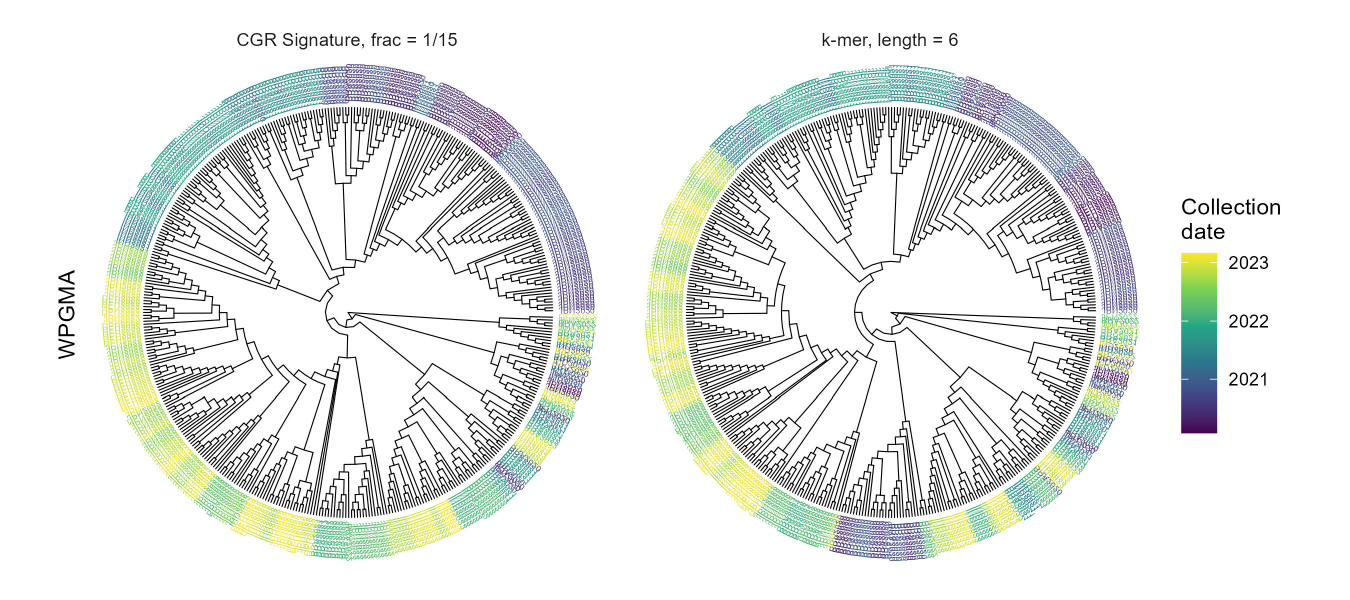}
\caption{Trees built on 434 coronavirus sequences (with an average sequence length of 29,784.48 base-pairs) collected between 2020 and early 2023. The trees built under WPGMA have been vertically inverted to facilitate comparison with the trees built under NJ by color.}
\label{fig:coronavirus}
\end{figure}

Plots of the trees for 434 SARS-CoV-2 sequences built using shape signature are provided on the left and plots of trees built under k-mer distance are provided on the right of Figure \ref{fig:coronavirus}. The colors correspond to the date the sequence was collected. Our expectation is that sequences collected during similar time periods should cluster together, presumably dominated by the most virulent strains of the time period. We also provide results under two different agglomeration methods that are representative of general trends observed under alternative agglomeration methods and discuss the most notable differences.\footnote{The CGR distance calculation took under 3 minutes to run and the k-mer distances took a few hours. However, it is not always the case that our method is faster; the speed of our method is largely a function of sequence length while the speed of the k-mer distance depends heavily on the parameter setting for the k-mer length. We expect our method to be slower for sufficiently long sequences.}

\FloatBarrier
\FloatBarrier

Overall, across all methods, we observe decent temporal clustering of sequences, particularly in the earlier data, presumably before the virus had diversified. However, we do see greater heterogeneity in clusters in later time-points. The heterogeneously-colored clusters highlight where both the k-mer approach and shape signature approach perform similarly poorly. 

However, we can observe the shape signature outperforming the k-mer distance in a few areas. In the top row of Figure \ref{fig:coronavirus}, under NJ agglomeration, the k-mer distances result in a set of 2022 sequences clustered within the set of much earlier sequences (shown as a patch of green within the purple cluster). Under WPGMA, the k-mer methods separate the early sequences into two different clusters (seen as the two sets of purple clusters). These sequences are likely misclustered due to not being temporally contiguous with adjacent sequences on the trees and at least one of these misclusterings always occurred under k-mer distance regardless of the clustering method chosen. The shape signature method never clustered the set of 2022 sequences within the earlier sequences. And any misclusterings of the early sequences was always substantially less severe, always retaining greater temporal proximity to adjacent sequences in the tree, or not observed in the shape signature method. 

We provide these trees for demonstrative purposes, but the point of our method is primarily the distances generated; the phylogenetic trees are simply an additional bonus feature by which our distance metrics may be used. Good distance measures should result in good phylogenetic trees, but a ``good" tree outcome does not guarantee that the distances are good. The tree construction process can mask or compensate for flaws in the distance data, as demonstrated by the differing topologies in Figure \ref{fig:coronavirus}. The final topology of these trees is dependent on not just the distances but also the type of agglomeration method used and on the set of organisms included in the distance matrix because the inclusion or exclusion of any one organism can change the topology of the tree. With such variability in the trees generated, one could conceivably permute their choice of agglomeration methods and organisms to generate seemingly ``good” clusters where ``good” is a subjective assessment of whether the clusters are consistent with expectations. 

Many alignment-free approaches prioritize speed and scalability over a more rigorous examination of the distance metrics and default to generating large trees as the primary validation method. We contend that massive, opaque trees should not be the standard by which we assess alignment-free methods and that there is space in this research area to do better and bridge the gap in insights provided by alignment-based methods and alignment-free methods. Unless strong ground-truth information is known, the building of such trees is arguably an exercise in confirmation bias. 

Our criticisms of the literature on alignment-free methods has been gently referenced throughout our entire paper. Those criticisms are that: 1) the choice of phylogenetic trees are subject to cherry-picking behavior by researchers; 2) many methods consistently show poor performance in the presence of unequal sequence lengths; and 3) alignment-free methods make minimal attempts to reflect the evolutionary distances and/or correlate alignment-free distance with true evolutionary distance. Such methods often perform well only under contrived scenarios. We make no claims that our method is uniformly superior for all sequences and all circumstances, but this paper is an attempt to address these criticisms and produce alignment-free distances that offer greater insight into the causes of sequence divergence. 

\section{Discussion}

Shape-based methods generate distances between sequences that correspond well to the number of observable separating mutation events and this feature is insensitive to parameter selection. Despite the volume intersection method producing pairwise distances that are broadly similar to k-mer distances for the $\beta$-globin gene and the MT-NADH gene, results from the synthetic sequences show that the shape-based methods produced distances that much more accurately reflect the number of mutation events between a parent and child sequence than both of the baseline reference methods. And these distances produced consistently decent quality downstream analyses. 

Shape-based methods do not disproportionately inflate the differences between sequences due to deletions versus substitutions and they remain comparable even after 70\% of the sequence has been mutated. They also perform well in the face of contiguous deletions, insertions, or transpositions that are likely due to single mutation events because they do not overly penalizing those types of mutations as though they were the result of multiple events.

For the real DNA sequences examined, the shape-based methods produced phylogenetic trees that are comparable to those built under an alignment-based method. The results are also mostly consistent with expectations given knowledge of taxonomic relationships between the selected organisms. Another feature of the shape-based methods is that the final phylogenetic trees are relatively invariant to parameter selection. Our explorations of the various parameter settings shows that even rough characterizations of a DNA sequence through 3D CGR can perform well. 

At every possible opportunity, we've made modeling choices to the benefit of the reference methods, Clustal Omega with p-distances and k-mer distances. The parameterization for the k-mer distance was set such that it would produce phylogenetic trees closest to that of the alignment-based method. Yet our method meets the performance of those methods and often outperforms them under alternative agglomeration methods. When looking at pairwise distances for our synthetic sequences, we observe that shape-based methods more closely reflect the number of observable mutation events. Increasing the resolution of the shape-based parameters (decreasing the bandwidth for the volume intersection and increasing the number of bins for the shape signature) tended to improve the estimated pairwise distances; while increasing the k-mer length did also improve its performance, it never matched the performance observed in the shape-based methods. 

With regards to the performance of our method on outside datasets, we also tested the shape signature method on 25 fish mitochondrial genomes provided by the AFproject \citep{AFproject} to compare against 105 other methods submitted to the project. The fish genomes were a minimum of 16,441 nucleotides and a maximum of 17,045. With the bin count set to 1/15th of the average sequence length, the method ranked 4th based on the Robinson-Fould metric; with the bin count set to 5 times the average sequence length, the method ranked 2nd (behind 32 methods tied for first place) and tied with 43 other methods. This improvement in ranking as the number of bins increases (up to a certain limit) is consistent with our expectations that increased resolution can improve distances. These results are promising in light of the relatively simple approach. We recover taxonomic information about the fish sequences from MitoFish, the Mitochondrial Genome Database of Fish \citep{Mitofish2013}, \cite{Mitofish2018}, \cite{Mitofish2023}), and present the results in the Appendix. 

However, the shape-based methods are not without their drawbacks. While, on average, an increase in the number of mutations tends to correspond to an increase in distance, an increase in distance is not guaranteed for every additional mutation event. And it is conceivable for two different sequences to have very similar, if not identical, signatures if too few bins are used for shape histograms. 

For the shape signature method, it is possible to increase the number of bins to capture more granular difference between sequences; but one has to be cognizant of creating shape signatures that are greater in size than the actual sequence itself. We have set the analysis parameters of this paper such that the total size of the shape signature representation is roughly equal to the average sequence length so that the space needed to store the signature is similar to that of the sequence. However, a more compressed representation may be desirable. We also noted that the improvement from increasing the resolution of the parameter settings (greater bin counts for shape signatures and smaller bandwidths for the volume intersection method) did eventually plateau and never fully reached the ground-truth estimates; increasing the resolution by an excessive amount also inflated distances to the detriment of phylogenies built by causing the overlap between sequences to approach zero. 

And, unfortunately, a significant drawback of the volume intersection method is that it can be too computationally intensive for practical applications. Since the volume of each sphere and their intersections is determined by a lengthy sampling process, even short sequences can take a very long time to run. Substantially longer sequences would become intractable. While we can decrease the bandwidth of the volume intersection method without any cost, the current limiting factor for the volume intersection method is sequence length, which for DNA sequences will always be quite long. 

Future work may find alternative features that generate histograms that better capture changes in sequences. We've demonstrated our approach on a tetrahedron, but the vertices themselves may also be adjusted to better capture desired characteristics in a distance calculation. For example, placing the purines in close proximity to each other in the space, but far from the pyrimidines would effectively penalize/increase the distance of transversion substitutions over transition substitutions as those mutations are rarer to aid in detecting selection or non-neutral evolutionary pressures. In addition to changing the vertices of the 3D CGR, we may also modify (or potentially drop some) features in the shape signature such that each type of substitution has a custom effect on the distance metric similar to molecular evolutionary models for DNA distance such as Kimura's 3-parameter distance \citep{Kimura1981-yk} or Tamura and Nei's 1993 model \citep{Tamura1993-ee}, which typically assign greater distance to transition mutations to correct for likely multiple substitutions at the same site over time. 

Alternative distance measures may still be considered such as an earth mover's distance for the histograms built from sequences of equal length. Overlapping kernel density estimates may also be used in place of histograms to better approximate the feature distributions. Although, currently, the volume intersection method is a computationally intensive task, we may be able to train a model that can approximate the volume intersection and substantially reduce the run-time of the volume intersection method. 

Ultimately, the shape-based methods are advantageous in that they: 1) do not require aligning the sequences, 2) retain much of the accuracy of alignment-based methods when building phylogenetic trees while reflecting the number of deletion and substitution events and, in the case of shape signatures, 3) can be computationally inexpensive.

\section*{Acknowledgement}

We thank Gordon Honerkamp-Smith and Tyler McCleery for helpful conversations early in the paper's conception and for Gordon's assistance in proofreading the manuscript.

\section*{Declaration of Generative AI and AI-assisted technologies in the writing process}

During the preparation of this work the author(s) submitted elements of the introduction to ChatGPT to edit for clarity. After using this tool/service, the author(s) reviewed and revised the content as needed and take(s) full responsibility for the content of the publication.

\section*{Conflict of interest}
Stephanie Young is a statistician at Johnson \& Johnson and received educational reimbursement from the company as part of an employee benefit program. Johnson \& Johnson did not influence or endorse the research presented in this paper.  

\appendix

\section{Fish mtDNA Results}

\FloatBarrier
\begin{figure}[ht] 
\includegraphics[width=\linewidth]{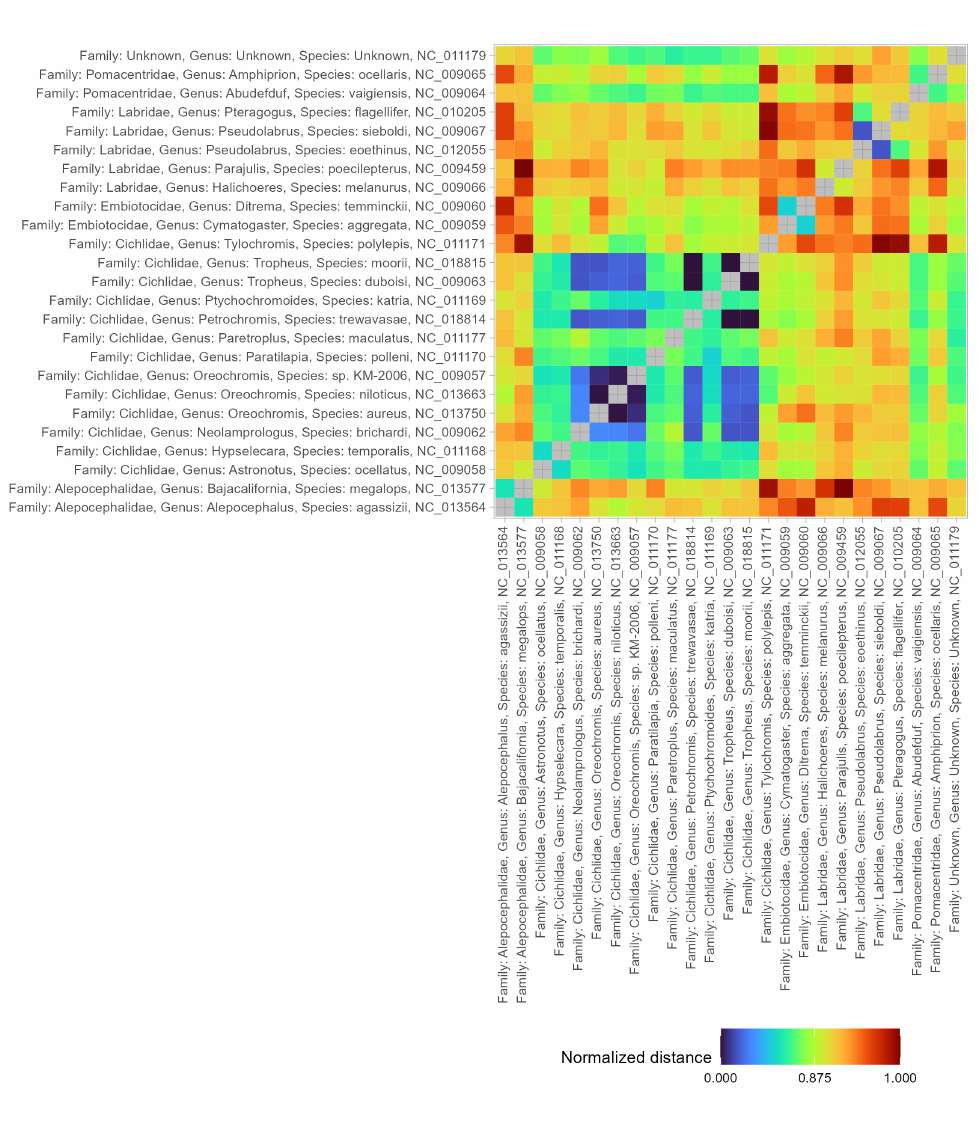}
\caption{Heatmap of distances for fish mtDNA with the number of bins for each histogram set to 5 times the average sequence length.}
\label{fig:fish-hm}
\end{figure}

\begin{figure}[ht] 
\includegraphics[width=1.1\linewidth]{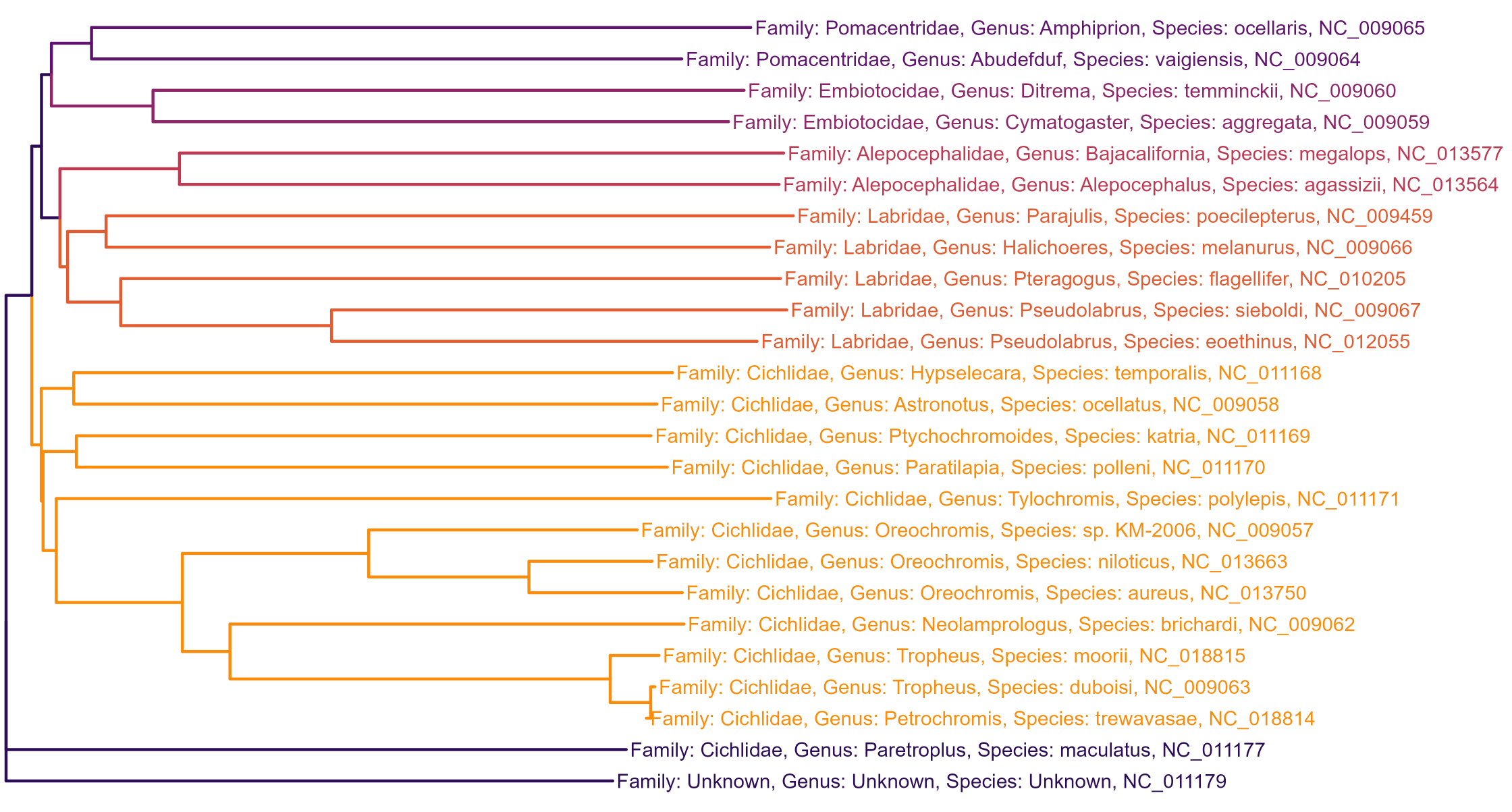}
\caption{Tree for fish mtDNA built with the number of bins for each histogram set to 5 times the average sequence length.}
\label{fig:fish-tree}
\end{figure}
\FloatBarrier

\FloatBarrier
\begin{figure}[ht] 
\includegraphics[width=\linewidth]{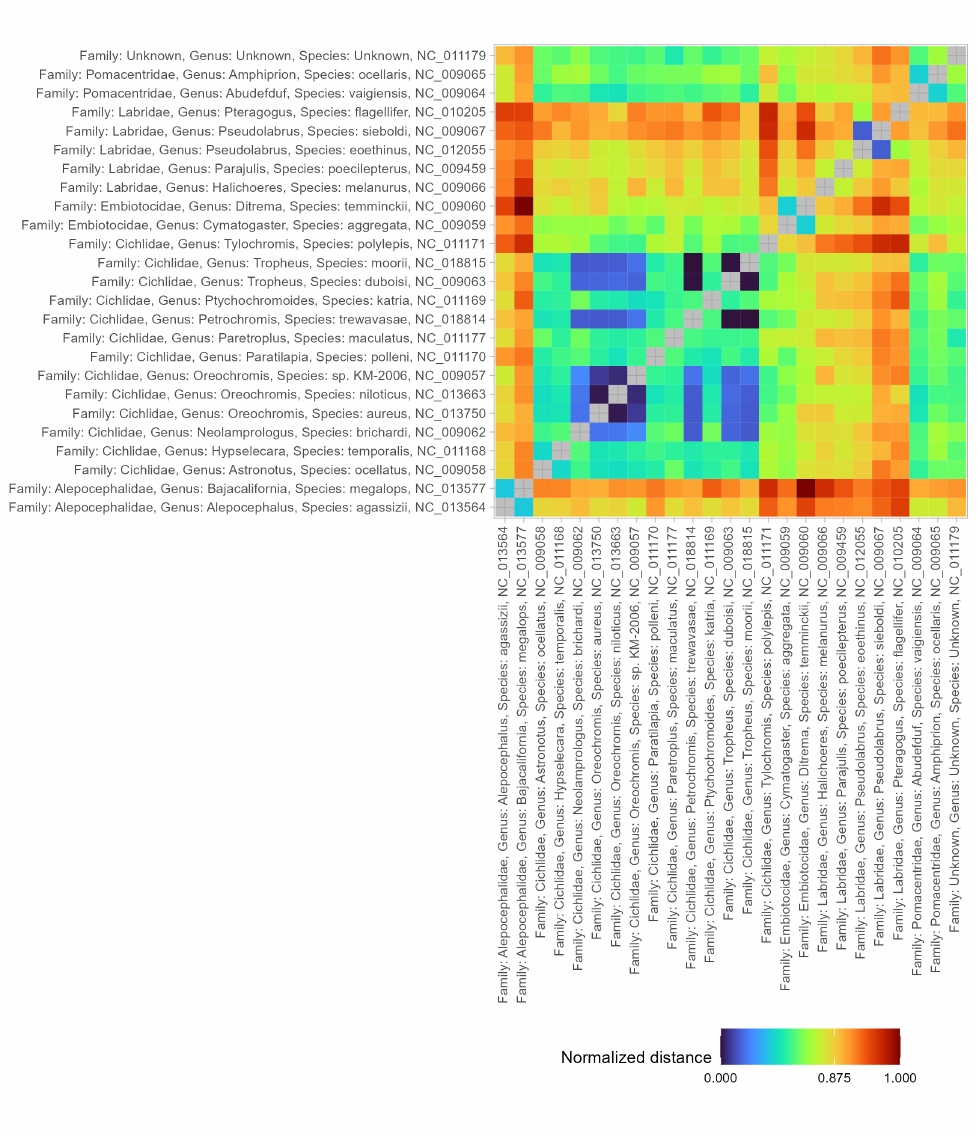}
\caption{Heatmap of distances for fish mtDNA using volume intersection (number of total points in estimation process capped at 300,000).}
\label{fig:fish-vi-hm}
\end{figure}

\begin{figure}[ht] 
\includegraphics[width=1.1\linewidth]{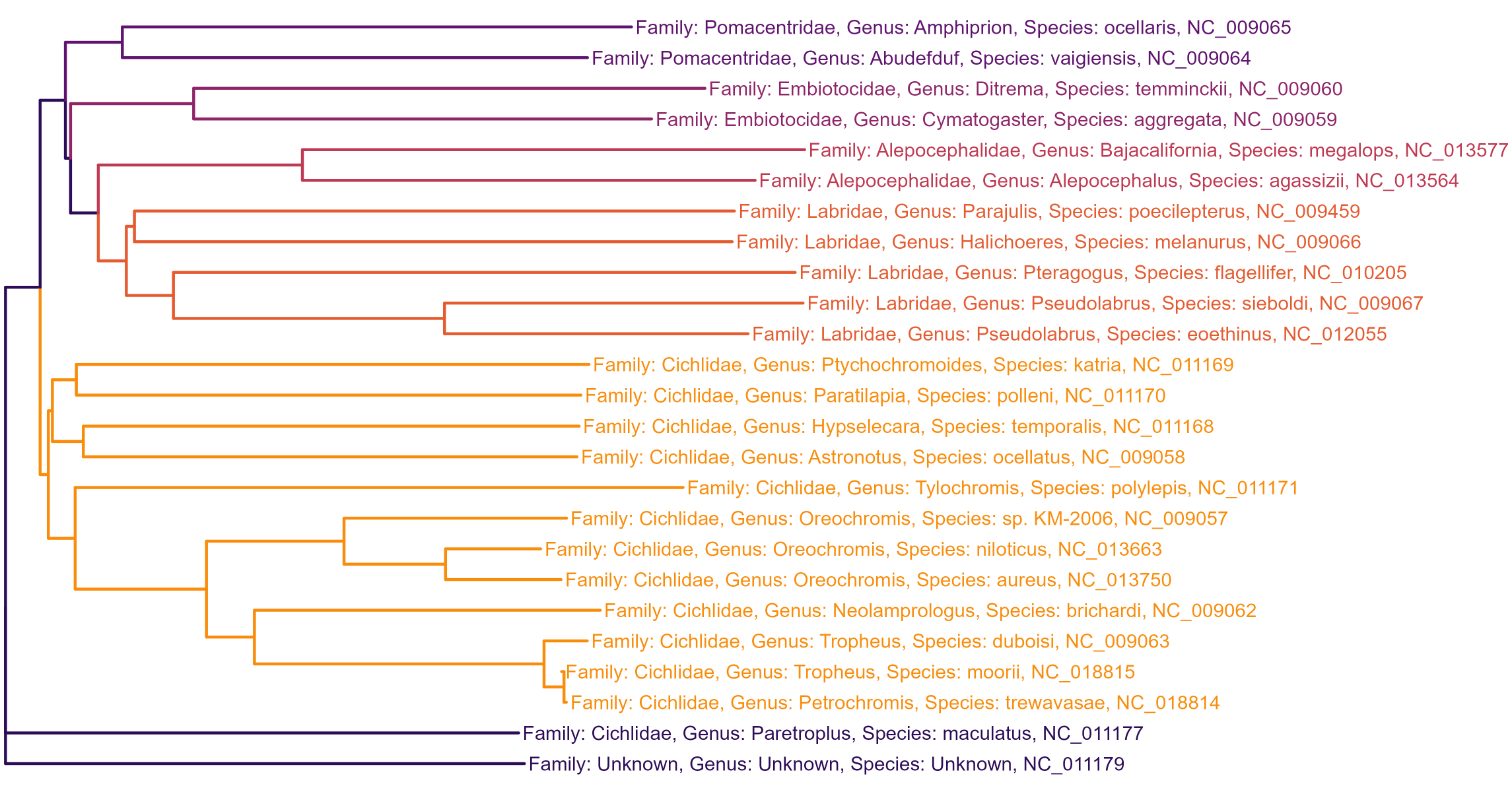}
\caption{Tree for fish mtDNA built with volume intersection of 3D CGR.}
\label{fig:fish-vi-tree}
\end{figure}
\FloatBarrier


\bibliographystyle{elsarticle-harv} 
\bibliography{thbib.bib}

\end{document}